\hfuzz 2pt
\font\titlefont=cmbx10 scaled\magstep1
\magnification=\magstep1

\null
\vskip 1.5cm
\centerline{\titlefont COMPLETELY POSITIVE DYNAMICS}
\smallskip
\centerline{\titlefont OF CORRELATED NEUTRAL KAONS}
\vskip 2.5cm
\centerline{\bf F. Benatti}
\smallskip
\centerline{Dipartimento di Fisica Teorica, Universit\`a di Trieste}
\centerline{Strada Costiera 11, 34014 Trieste, Italy}
\centerline{and}
\centerline{Istituto Nazionale di Fisica Nucleare, Sezione di 
Trieste}
\vskip 1cm
\centerline{\bf R. Floreanini}
\smallskip
\centerline{Istituto Nazionale di Fisica Nucleare, Sezione di 
Trieste}
\centerline{Dipartimento di Fisica Teorica, Universit\`a di Trieste}
\centerline{Strada Costiera 11, 34014 Trieste, Italy}
\vskip 2cm
\centerline{\bf Abstract}
\smallskip
\midinsert
\narrower\narrower\noindent
We study the behaviour of correlated neutral kaons 
produced in \hbox{$\phi$-meson} decays under the hypothesis that the 
quantum mechanical time-evolution is completely positive.
We show that planned experiments at $\phi$-factories could give
precise bounds on the phenomenological parameters of the model.
\endinsert
\bigskip
\vfil\eject

{\bf 1. INTRODUCTION}
\medskip

Recently, it has been proposed to describe the neutral kaon system using
effective dynamics not of the standard Weisskopf-Wigner type.
These non-standard time-evolutions transform pure states 
into mixed ones and lead to $CP$ and $CPT$ violating effects [1-4, 5].
The physical motivations behind such an approach are based on quantum gravity,
that predicts loss of coherence at the Planck's length 
due to fluctuations of the gravitational field [6, 7]. 
Remarkably, the neutral kaon system provides an experimental testing
ground for this hypothesis [8] giving bounds to the
phenomenological parameters of the theory.

The non-standard quantum dynamics discussed in [1-4, 5] is based on 
a specific evolution map that acts on the kaon density matrix $\rho$,
$\rho\mapsto\rho(t)$. This map has the property of being (simply) positive,
since it preserves the positivity of the eigenvalues of the density matrix
$\rho(t)$ for all times $t\geq 0$; it further fulfills other basic physical
requirements: it decreases the trace of $\rho(t)$, 
${\rm d}/{\rm dt}\big[\hbox{Tr}\rho(t)\big]\leq0$, and 
increases the entropy, 
$-{\rm d}/{\rm dt}\big[\hbox{Tr}\rho(t)\log\rho(t)\big]\geq0$.
The non-unitarity of this new dynamics is not only due to the 
decay of the neutral kaons, as in the standard Weisskopf-Wigner 
phenomenological description, but also to the fact that 
initially pure kaon states become statistical mixtures.

A different, more general approach is however possible;
it treats decay systems as specific examples of open quantum systems.[9-11]
These systems can be modeled as being small
subsystems in interaction with suitable large environments.
The global evolution of the closed compound system is described by an
unitary map, while the reduced dynamics of the subsystem usually
develops dissipation and irreversibility.
Assuming a weak coupling between subsystem and
environment, the reduced time-evolution that results from eliminating
the environment degrees of freedom is free from non-linear feedback or
memory effects, thus  possessing very basic and fundamental properties, like
forward in time composition (semigroup property), probability conservation
and entropy increase. 
Moreover, the corresponding dynamical map $\gamma_t:\rho\mapsto\rho(t)$
has the additional property of being completely positive [12-15],
and not only simply positive.
This set of transformations forms a so-called dynamical semigroup. 

In the case of the neutral kaons,[16-19] the complete positivity of the time
evolution $\gamma_t$ amounts to the positivity of the natural extension of 
$\gamma_t$ to a linear transformation on the states of a larger system
consisting of the kaon system and another finite-level system of 
arbitrary dimension. 
As we have already remarked, simple positivity guarantees that the 
eigenvalues of any density
matrix for the neutral kaon system remain positive. Complete positivity 
is a stronger property in the sense that the same holds for 
density matrices of the compound system.

Although at first sight the requirement of complete positivity 
instead of the much milder simple positivity seems
a mere technical request, it has far-reaching consequences. 
In particular, when the additional finite-level system alluded above
is taken to be another kaon system (a physical situation typically
encountered in $\phi$-meson decays), the complete positivity of the
time-evolution $\gamma_t$ assures the absence of unphysical effects, like
the appearance of negative probabilities, that could occur for simple
positive dynamics [18].

In [16, 17] the decay of a single neutral kaon system was 
studied under the hypothesis of
a completely positive non-standard quantum dynamics and the 
phenomenological parameters specifying the theory were compared with
presently available experimental data [19].
Here, we extend that analysis to the case of
correlated neutral kaons under the assumption
that, after being produced in a $\phi$-meson decay, they evolve
independently under the completely positive quantum dynamics.
This analysis is motivated by the new experimental facilities that
will soon be available for the study of $\phi$-meson decays,
the so called $\phi$-factories.
The outcome of our work is that planned experiments 
at these facilities are likely to
put stringent bounds on the phenomenological parameters governing
the proposed models.

In Section 2 we describe briefly the dynamics of a single kaon under the
hypothesis of complete positivity. This dynamics is then applied in Section 3
to the analysis of the time evolution of correlated kaons, as produced
by the decay of a $\phi$-meson. 
As explained in Section 4, this allows a detailed
study of the time-evolution of various physical observables that will be
experimentally accessible at $\phi$-factories. Finally, Section 5 comprises
concluding discussions and comments, as well as additional considerations
concerning completely positive {\it vs} simply positive dynamics for
correlated kaons. 

\vskip 1cm

{\bf 2. SINGLE KAON COMPLETELY POSITIVE DYNAMICS}
\medskip

As usual, we shall model the evolution and decay of 
the $K^0$-$\overline{K^0}$ system
by means of a two-dimensional Hilbert space [20].
A convenient orthonormal basis in this space is given by the $CP$-eigenstates 
$|K_1\rangle$ and $|K_2\rangle$:
$$
|K_1\rangle={1\over\sqrt{2}}\Big[|K^0\rangle+|\overline{K^0}\rangle\Big]
\ ,\quad
|K_2\rangle={1\over\sqrt{2}}\Big[|K^0\rangle-|\overline{K^0}\rangle\Big]\ .
\eqno(2.1)
$$
Single kaon states will be described by density
matrices $\rho$, {\it i.e.} by hermitian $2\times 2$ matrices 
with positive eigenvalues and unit trace.
With respect to the basis (2.1), one can write:
$$
\rho=\left(\matrix{
\rho_1&\rho_3\cr
\rho_4&\rho_2}\right)\ , \eqno(2.2)
$$
where $\rho_4\equiv\rho_3^*$, and $*$ signifies complex conjugation.

As explained in the Introduction, our analysis is based on the assumption
that the evolution in time of a given initial kaon state $\rho$ be completely
positive. The corresponding evolution map,
$\rho\mapsto\rho(t)\equiv\gamma_t[\rho]$, is then generated by an equation of
the following form:
$$
{\partial\rho(t)\over\partial t}=-iH\, \rho(t)+i\rho(t)\, H^\dagger 
+L[\rho] .\eqno(2.3)
$$
The first two pieces in the r.h.s. constitute the standard Weisskopf-Wigner
contribution, while $L$ is a linear map that is fully determined 
by the request of complete positivity (and trace conservation):
$$
L[\rho]=-{1\over2}\sum_j\Big(A^\dagger_jA_j\,\rho +
\rho\, A^\dagger_jA_j\Big)\ 
+\sum_j A_j\,\rho\, A^\dagger_j\ . \eqno(2.4)
$$
The operators $A_j$ must be such that $\sum_j A^\dagger_j A_j$ is a
well-defined $2\times 2$ matrix; further, to assure
entropy increase, the $A_j$ will be taken to be hermitian.

In absence of $L[\rho]$, pure states ({\it i.e.} states of the form 
$|\psi\rangle\langle\psi|$) would be transformed into pure states, 
in spite of the fact that probability is not conserved.
However, this is due to the presence of a non-hermitian part in the effective 
hamiltonian $H$ and not to a mixing-enhancing mechanism
that makes $\rho(t)$ less ordered in time. Instead, loss of quantum 
coherence shows up when the extra piece $L[\rho]$ is also present:
it produces dissipation and possible transitions from pure states to
mixed states.

The effective hamiltonian $H$ (the Weisskopf-Wigner hamiltonian)
includes a nonhermitian part, that characterizes 
the natural width of the states:
$$
H=M-{i\over 2}{\mit\Gamma}\ ,\eqno(2.5)
$$ 
with $M$ and $\mit\Gamma$ hermitian $2\times 2$ matrices.
The entries of these matrices can be expressed in terms of 
the four real parameters, $m_S$, $\gamma_S$ and $m_L$, $\gamma_L$
characterizing the eigenvalues of $H$: 
$$
\lambda_S=m_S-{i\over 2}\gamma_S\ ,\quad
\lambda_L=m_L-{i\over 2}\gamma_L\ ,\eqno(2.6)
$$
and the two complex parameters $\epsilon_S$, $\epsilon_L$, appearing in the
eigenstates of $H$, 
$$
\eqalign{
|K_S\rangle&=N_S\Big(|K_1\rangle\ +\ \epsilon_S|K_2\rangle\Big)\cr
|K_L\rangle&=N_L\Big(\epsilon_L|K_1\rangle\ +\ |K_2\rangle\Big)\ ,\cr
}\eqno(2.7)
$$
with $N_S=(1+|\epsilon_S|^2)^{-1/2}$ and $N_L=(1+|\epsilon_L|^2)^{-1/2}$ 
normalization factors.
It proves convenient to use also the following
positive combinations:
$$
\Delta\Gamma=\gamma_S-\gamma_L\ ,\qquad
\Delta m=m_L-m_S\ ,\eqno(2.8)
$$
corresponding to the differences between decay widths and masses of the
states $K_S$ and $K_L$, as well as of the complex quantities:
$$
\Gamma_\pm=\Gamma\pm i \Delta m\ ,\qquad 
\Delta\Gamma_\pm=\Delta\Gamma\pm 2i\Delta m\eqno(2.9)
$$
with $\Gamma=(\gamma_S+\gamma_L)/2$.

The explicit form of the piece $L[\rho]$ can be most simply given by 
expanding $\rho$ in terms of Pauli matrices $\sigma_i$ and the identity
$\sigma_0$: $\rho=\rho_\mu\, \sigma_\mu$, $\mu=\,0$, 1, 2, 3.
In this way, the map $L[\rho]$ can be represented by
a symmetric $4\times 4$ matrix $\big[L_{\mu\nu}\big]$, 
acting on the column vector with components $(\rho_0,\rho_1,\rho_2,\rho_3)$.
It can be parametrized by the six real 
constants $a$, $b$, $c$, $\alpha$, $\beta$, and $\gamma$: [16, 17]
$$
\big[L_{\mu\nu}\big]=-2\left(\matrix{0&0&0&0\cr
                                     0&a&b&c\cr
                                     0&b&\alpha&\beta\cr
                                     0&c&\beta&\gamma\cr}\right)
\ ,\eqno(2.10)
$$
with $a$, $\alpha$ and $\gamma$ non-negative.
These parameters are not all independent; to assure the complete positivity
of the time-evolution $\rho\rightarrow\rho(t)$, they have to satisfy
the following inequalities:
$$
\eqalign{&a\leq \alpha+\gamma\ ,\phantom{\big(\beta^2\big)^2}\cr
         &\alpha\leq a+\gamma\ ,\phantom{\big(\beta^2\big)^2}\cr
         &\gamma\leq a+\alpha\ ,\phantom{\big(\beta^2\big)^2}\cr}\quad
\eqalign{&4b^2\leq \gamma^2-\big(a-\alpha\big)^2\ ,\cr
         &4c^2\leq \alpha^2-\big(a-\gamma\big)^2\ ,\cr
         &4\beta^2\leq a^2-\big(\alpha-\gamma\big)^2\ .\cr}
\eqno(2.11)
$$

Physical observables of the neutral kaon system can be obtained from the density
matrix $\rho(t)$ obeying (2.3) by taking its trace with suitable hermitian
operators. In order to study the time evolution of these observables, one has to
solve the equation (2.3) for a given initial state $\rho$, {\it i.e.}
compute the entries of the $4\times 4$ evolution matrix $M_{ij}(t)$,
which in the basis (2.1), (2.2) gives the entries of the density matrix
at time $t$:
$$
\rho_i(t)=\sum_{j=1}^4 M_{ij}(t)\, \rho_j\ ,\qquad i=1,2,3,4\ .
\eqno(2.12)
$$

According to plausible phenomenological considerations,[1-4, 5, 17] 
the parameters $a$, $b$, $c$, $\alpha$, $\beta$ and $\gamma$ can be assumed
to be small, of the same order of magnitude of $\epsilon_S \Delta\Gamma$ and
$\epsilon_L \Delta\Gamma$; the use of perturbation theory is therefore
justified. For most purposes, it is sufficient to consider contributions up
to the second order in these small parameters.
The expansion of $M_{ij}(t)$ within this approximation can be conveniently
organized as the sum of three contributions:[17]
$$
\rho_i(t)\simeq\sum_{j=1}^4
\Big[ M^{(0)}_{ij}(t)\ +\ M^{(1)}_{ij}(t)\ + \
M^{(2)}_{ij}(t)\Big]\, \rho_j\ . \eqno(2.13)
$$
For further reference, we list below the explicit expressions of the
entries in the matrices $M^{(0)}$ and $M^{(1)}$. The contributions of 
$M^{(2)}$ can be found in [17], together with a detailed discussion of the
method used in their derivation. The matrix $M^{(0)}$ has only diagonal
non-vanishing entries:
$$
M^{(0)}_{11}(t)=e^{-\gamma_S t}\ ,\quad
M^{(0)}_{22}(t)=e^{-\gamma_L t}\ ,\quad
M^{(0)}_{33}(t)=e^{-\widetilde{\Gamma}_-t}\ ,\quad
M^{(0)}_{44}(t)=e^{-\widetilde{\Gamma}_+t}\ ,\eqno(2.14)
$$
where
$$
\widetilde{\Gamma}_\pm=\Gamma_\pm + A-\gamma
-2\, {\cal R}e\left[C(\epsilon_S^*-\epsilon_L)\right]
\mp i{|B|^2\over {2\Delta m}}\mp 8i\, \Delta m\,
\left|{C\over{\Delta\Gamma_+}}\right|^2\ ,
\eqno(2.15)
$$
and $A$, $B$, $C$ are the following convenient combinations of the parameters in
(2.10):
$$
A=\alpha+a\ ,\quad B=\alpha-a+2ib\ ,\quad C=c+i\beta\ .\eqno(2.16)
$$
The entries of $M^{(1)}$ take instead the following explicit expression:
$$
\eqalign{
&M^{(1)}_{11}(t)=0 \phantom{\gamma\over\Delta\Gamma}\cr
&M^{(1)}_{13}(t)=e^*_L(-,-)\Big(
                  e^{-(\Gamma_- +A-\gamma) t}-e^{-\gamma_S t}\Big)\ \cr
&M^{(1)}_{21}(t)={\gamma\over\Delta\Gamma}\Big(
                  e^{-\gamma_L t}-e^{-\gamma_S t}\Big)\ \cr
&M^{(1)}_{23}(t)=f_S(-,-)\Big(e^{-(\Gamma_- +A-\gamma)  t}
                               -e^{-\gamma_L t}\Big)\ \cr
&M^{(1)}_{31}(t)=f^*_S(+,-)\Big(e^{-\gamma_S t}
                                 -e^{-(\Gamma_- +A-\gamma) t}\Big)\ \ \cr
&M^{(1)}_{33}(t)=0 \phantom{iB\over2\Delta m}\cr
&M^{(1)}_{41}(t)=f_S(+,-)\Big(e^{-\gamma_S t}
                               -e^{-(\Gamma_+ +A-\gamma) t}\Big)\ \cr
&M^{(1)}_{43}(t)={iB^*\over2\Delta m}\,
                  e^{-(A-\gamma)t}\Big(e^{-\Gamma_+ t}
                       -e^{-\Gamma_- t}\Big)\ }
\eqalign{
&M^{(1)}_{12}(t)={\gamma\over\Delta\Gamma}
                  \Big(e^{-\gamma_L t}-e^{-\gamma_S t}\Big)\cr
&M^{(1)}_{14}(t)=e_L(-,-)\Big(e^{-(\Gamma_+ +A-\gamma)  t}
                 -e^{-\gamma_S t}\Big)\cr
&M^{(1)}_{22}(t)=0 \phantom{\gamma\over\Delta\Gamma}\cr
&M^{(1)}_{24}(t)=f^*_S(-,-)\Big(e^{-(\Gamma_+ +A-\gamma)  t}
                                 -e^{-\gamma_L t}\Big)\cr
&M^{(1)}_{32}(t)=e_L(+,-)\Big(e^{-\gamma_L t}
-e^{-(\Gamma_- +A-\gamma) t}\Big)\cr
&M^{(1)}_{34}(t)={iB\over2\Delta m}\, e^{-(A-\gamma)t}
                   \Big(e^{-\Gamma_+ t}-e^{-\Gamma_- t}\Big)\cr
&M^{(1)}_{42}(t)=e^*_L(+,-)\Big(e^{-\gamma_L t}
                                 -e^{-(\Gamma_+ +A-\gamma) t}\Big)\cr
&M^{(1)}_{44}(t)=0\ ,\phantom{iB\over2\Delta m}}
\eqno(2.17)
$$
where we have introduced the shorthand notations:
$$
\eqalign{
&e_{S,L}(+,\pm)=\epsilon_{S,L}+{2C^*\over{\Delta\Gamma_\pm}}\ ,\cr
&e_{S,L}(-,\pm)=\epsilon_{S,L}-{2C^*\over{\Delta\Gamma_\pm}}\ ,}
\qquad
\eqalign{
&f_{S,L}(+,\pm)=\epsilon_{S,L}+{2C\over{\Delta\Gamma_\pm}}\ ,\cr
&f_{S,L}(-,\pm)=\epsilon_{S,L}-{2C\over{\Delta\Gamma_\pm}}\ .}
\eqno(2.18)
$$
Notice that in presenting the above expressions for the entries of 
$M^{(0)}$ and $M^{(1)}$, we have reconstructed the exponential dependences
out of first and second order correction terms which are linear and quadratic
in time. To a given order in the perturbative
expansion, this can always be done provided one suitably redefines
$\Gamma_\pm$ and $\Delta\Gamma_\pm$ in (2.9) in such a way that $\gamma_S$,
$\gamma_L$ and $\Delta m$ can be directly identified with the widths
and mass-difference of the $K_S$ and $K_L$ physical states (for more details,
see the discussion in the Appendix of [17]).

In [17, 19] the approximate solution (2.13) to the evolution equation
(2.3) has been used to compute the time dependence of many physical observables
of the single kaon system relevant to the experiment. In the next section, we
shall apply it to the study of the time evolution of correlated systems of
two neutral kaons.

\vskip 1cm

{\bf 3. COMPLETELY POSITIVE DYNAMICS OF CORRELATED KAONS}
\medskip

The time evolution of a system of two correlated neutral kaons can be discussed
using the results on the dynamics of a single kaon of the previous section.
We shall limit the analysis to the study of entangled $K^0$-$\overline{K^0}$
states coming from the decay of a $\phi$-meson. This is of great experimental
relevance in view of the possibilities offered by $\phi$-factories.

Since the $\phi$-meson has spin 1, its decay into two spinless bosons produces
an antisymmetric spatial state. In the $\phi$ rest frame, the two neutral kaons
are produced flying apart with opposite momenta; in the basis $|K_1\rangle$,
$|K_2\rangle$, the resulting state can be described by:
$$
|\psi_A\rangle= {1\over\sqrt2}\Big(|K_1,-p\rangle \otimes  |K_2,p\rangle -
|K_2,-p\rangle \otimes  |K_1,p\rangle\Big)\ .\eqno(3.1)
$$
The corresponding density operator $\rho_A$ is a $4\times 4$ matrix
that can be conveniently written in terms of single kaon projectors:
$$
P_1\equiv |K_1\rangle\langle K_1|=
    \left(\matrix{1 & 0\cr 0 & 0\cr}\right)\ ,\qquad
P_2\equiv |K_2\rangle\langle K_2|=
    \left(\matrix{0 & 0\cr 0 & 1\cr}\right)\ ,\eqno(3.2a)
$$
and the off-diagonal operators
$$
P_3\equiv |K_1\rangle\langle K_2|=
    \left(\matrix{0 & 1\cr 0 & 0\cr}\right)\ ,\qquad
P_4\equiv |K_2\rangle\langle K_1|=
    \left(\matrix{0 & 0\cr 1 & 0\cr}\right)\ .\eqno(3.2b)
$$
Explicitly, one finds:
$$
\rho_A={1\over 2}\Big[P_1\otimes P_2\ +\ P_2\otimes P_1\ -\ 
P_3\otimes P_4\ -\ P_4\otimes P_3\Big]\ .\eqno(3.3)
$$

In studying the dynamics of correlated kaons, we shall assume that, once
produced in a $\phi$ decay, the kaons evolve in time each according to the
completely positive map $\gamma_t$ described in the previous section.
This assures that the resulting evolution map 
$\rho_A\mapsto \Gamma_t[\rho_A]$ is
completely positive and of semigroup type; further, this dynamics
is independent from the particular situation under study and can be easily
generalized to systems containing more than two particles.
Although other possibilities are conceivable, this choice is the most
natural one. In fact, it is very hard to produce dynamical maps $\Gamma_t$ 
for a system of two particles not in factorized form
without violating very basic physical principles. Indeed, if one requires that 
after tracing over the degrees of freedom of one particle, the resulting
dynamics for the remaining one be $a)$ completely positive, 
$b)$ of semigroup type and $c)$ independent from 
the initial state of the first particle, than the only
natural possibility is: $\Gamma_t=\gamma_t \otimes \gamma_t$.

With this choice, the density matrix that describes a situation in which
the first kaon has evolved up to proper time $\tau_1$ 
and the second up to proper time $\tau_2$ is given by:
$$
\eqalign{
\rho_A(\tau_1&,\tau_2)\equiv
\big(\gamma_{\tau_1}\otimes\gamma_{\tau_2}\big)\big[\rho_A\big]\cr
=&{1\over 2}\Big[P_1(\tau_1)\otimes P_2(\tau_2)\ 
+\ P_2(\tau_1)\otimes P_1(\tau_2)\ 
- P_3(\tau_1)\otimes P_4(\tau_2)-P_4(\tau_1)\otimes P_3(\tau_2)\Big]\ ,}
\eqno(3.4)
$$
where $P_i(\tau_1)$ and $P_i(\tau_2)$, $i=1,2,3,4$, represent the evolution
according to (2.3) of the initial operators (3.2), up to the time $\tau_1$
and $\tau_2$, respectively.

As already noticed, the system of the two neutral kaons that is produced in the
decay of a $\phi$-meson is highly correlated and therefore constitutes a unique
setup for studying phenomena involving loss of quantum coherence. The typical
observables that can be studied in such quantum interferometer are double
decay rates, {\it i.e.} the probabilities that a kaon decays
into a final state $f_1$ at proper time $\tau_1$, while the other kaon 
decays into the final state $f_2$ at proper time $\tau_2$.[21-25]
Such correlations will be denoted by ${\cal G}(f_1,\tau_1;f_2,\tau_2)$.
Using (3.4), one explicitly finds:
$$
\eqalign{
{\cal G}(f_1,\tau_1;& f_2,\tau_2)\equiv 
\hbox{Tr}\Big[\Big({\cal O}_{f_1}\otimes{\cal O}_{f_2}\Big) 
\rho_A(\tau_1,\tau_2)\Big]\cr
=&{1\over 2}\Big[
\hbox{Tr}\big\{P_1(\tau_1)\,{\cal O}_{f_1}\big\}\
\hbox{Tr}\big\{P_2(\tau_2)\,{\cal O}_{f_2}\big\}\ +\
\hbox{Tr}\big\{P_2(\tau_1)\,{\cal O}_{f_1}\big\}\ 
\hbox{Tr}\big\{P_1(\tau_2)\,{\cal O}_{f_2}\big\}\cr
-&\hbox{Tr}\big\{P_3(\tau_1)\,{\cal O}_{f_1}\big\}\
\hbox{Tr}\big\{P_4(\tau_2)\,{\cal O}_{f_2}\big\}\ -\
\hbox{Tr}\big\{P_4(\tau_1)\,{\cal O}_{f_1}\big\}\
\hbox{Tr}\big\{P_3(\tau_2)\,{\cal O}_{f_2}\big\}\Big]\ .
}\eqno(3.5)
$$
In this formula, ${\cal O}_{f_1}$ and ${\cal O}_{f_2}$ represent the
$2\times 2$ hermitian matrices describing the decay of a single kaon into the
final states $f_1$ and $f_2$, respectively.
Any operator $\cal O$ of such type can be written in the basis (2.1) using the
the matrices $P_i$, $i=1,2,3,4$ of (3.2):
$$
{\cal O}=\sum_{i=1}^4 {\cal O}^i\ P_i\ . \eqno(3.6)
$$
Therefore, the observables (3.5) can be rewritten as
$$
{\cal G}(f_1,\tau_1; f_2,\tau_2)=
{1\over 2}\sum_{i,j=1}^4 {\cal O}_{f_1}^i\, {\cal O}_{f_2}^j\ 
{\cal P}_{ij}(\tau_1,\tau_2)\ ,\eqno(3.7)
$$
where
$$
\eqalign{
{\cal P}_{ij}(\tau_1,\tau_2)
&=\hbox{Tr}\big\{P_i\, P_1(\tau_1)\big\}\
\hbox{Tr}\big\{P_j\, P_2(\tau_2)\big\}\ +\
\hbox{Tr}\big\{P_i\, P_2(\tau_1)\big\}\ 
\hbox{Tr}\big\{P_j\, P_1(\tau_2)\big\}\cr
&\qquad
-\hbox{Tr}\big\{P_i\, P_3(\tau_1)\big\}\
\hbox{Tr}\big\{P_j\, P_4(\tau_2)\big\}\ -\
\hbox{Tr}\big\{P_i\, P_4(\tau_1)\big\}\
\hbox{Tr}\big\{P_j\, P_3(\tau_2)\big\}\ .
}\eqno(3.8)
$$

These ``elementary'' double probabilities ${\cal P}_{ij}$ can be computed
in perturbation theory using the solution to the evolution equation (2.3)
discussed in Section 2. Expansions for ${\cal P}_{ij}$ up to second order
in the small parameters will suffice to produce accurate enough expressions for
the correlations $\cal G$ that will be examined in the following.
The explicit expressions of the various probabilities ${\cal P}_{ij}$
to this degree of accuracy are collected in Appendix A. One finds that
${\cal P}_{12}$, ${\cal P}_{21}$, ${\cal P}_{34}$ and ${\cal P}_{43}$
possess zero-th and second order contributions in the small parameters, 
but not first order terms. All the other probabilities ${\cal P}_{ij}$ 
have only first and second order contributions.

As we shall see in the next section, the completely positive dynamics
(3.4) produces results for the double probabilities 
${\cal G}(f_1,\tau_1; f_2,\tau_2)$ that substantially differ from the
ones obtained using standard quantum mechanics. The most striking difference
arises in considering correlations at equal proper times $\tau_1=\tau_2=\tau$,
with the same final state $f_1=f_2=f$. Due to the antisymmetry of the initial
state $\rho_A$ in (3.3), ordinary quantum mechanics predicts a vanishing result
for ${\cal G}(f,\tau; f,\tau)$. This is in general not the case for the
completely positive dynamics discussed here. The equal-time correlations
${\cal G}(f,\tau; f,\tau)$ are therefore very sensible to the non-standard
parameters $a$, $b$, $c$, $\alpha$, $\beta$ and $\gamma$.
A direct experimental determination of the probabilities
${\cal G}(f,\tau; f,\tau)$ is clearly problematic; however, these correlations
can be extrapolated from the measured double-time ones,
${\cal G}(f,\tau_1; f,\tau_2)$, when the interval $\tau_1-\tau_2$ becomes small.
Therefore, the correlations ${\cal G}(f,\tau; f,\tau)$ can be very useful
for obtaining experimental estimates for the parameters
$a$, $b$, $c$, $\alpha$, $\beta$ and $\gamma$.

\vskip 1cm

{\bf 4. OBSERVABLES}
\medskip

A high-luminosity $\phi$-factory has been recognized as one of the best
experimental setup for studying $CP$ and $CPT$ violating effects.[21-25]
The machine,
producing $\phi$-mesons at high rate, behaves as a very accurate quantum
interferometer capable of revealing tiny effects and of measuring very
small quantities. Therefore, it is the natural experimental setup for
measuring probabilities involving double kaon decays of the type introduced
in the previous section. In the discussion that follows, we shall ignore
effects due to finite detector size and to background processes that produce
couple of $K^0$-$\overline{K^0}$ in a symmetric spin zero combinations.
We shall comment on some of these background effects in the final section.

In order to compute the correlations ${\cal G}(f_1,\tau_1; f_2,\tau_2)$,
one needs to give explicit expressions for the matrices ${\cal O}_f$
that describe the decay of a single kaon into the final state $f$.
Useful observables $\cal G$ are associated with the decays of the neutral
kaons into two and three pions and into semileptonic states. We shall be as
general as possible and use matrices ${\cal O}_f$ that encode possible
$CP$ and $CPT$ violating effects also in the decay amplitudes.

In the case of the $\pi^+\pi^-$ and $\pi^0\pi^0$ final states, using the basis
$K_1$, $K_2$ of (2.1), we shall label the corresponding decay amplitudes as:
$$
\eqalign{
&{\cal A}(K_1\rightarrow \pi^+\pi^-)=X_{+-}\ ,\cr
&{\cal A}(K_1\rightarrow \pi^0\pi^0)=X_{00}\ ,}\qquad\quad
\eqalign{
&{\cal A}(K_2\rightarrow \pi^+\pi^-)=Y_{+-}\ 
{\cal A}(K_1\rightarrow \pi^+\pi^-)\ ,\cr
&{\cal A}(K_2\rightarrow \pi^0\pi^0)=Y_{00}\ 
{\cal A}(K_1\rightarrow \pi^0\pi^0)\ .}
\eqno(4.1)
$$
The operators $\cal O$ that describe these final states take then the following
form [5, 19]:
$$
{\cal O}_{+-}=|X_{+-}|^2\ \left[\matrix{1&Y_{+-}\cr
                              Y_{+-}^*&|Y_{+-}|^2\cr}\right]\ ,\qquad
{\cal O}_{00}=|X_{00}|^2\ \left[\matrix{1&Y_{00}\cr
                              Y_{00}^*&|Y_{00}|^2\cr}\right]\ .
\eqno(4.2)
$$
Similar expressions hold for the decay into three pions, $\pi^+\pi^-\pi^0$
(with total isospin $I=1$ [20]) and $\pi^0\pi^0\pi^0$. 
Parametrizing the decay amplitudes as
$$
\eqalign{
&{\cal A}(K_2\rightarrow \pi^+\pi^-\pi^0)=X_{+-0}\ ,\cr
&{\cal A}(K_2\rightarrow \pi^0\pi^0\pi^0)=X_{000}\ ,}\qquad\quad
\eqalign{
&{\cal A}(K_1\rightarrow \pi^+\pi^-\pi^0)=Y_{+-0}\ 
{\cal A}(K_2\rightarrow \pi^+\pi^-\pi^0)\ ,\cr
&{\cal A}(K_1\rightarrow \pi^0\pi^0\pi^0)=Y_{000}\ 
{\cal A}(K_2\rightarrow \pi^0\pi^0\pi^0)\ ,}
\eqno(4.3)
$$
one finds [5, 19]
$$
{\cal O}_{+-0}=|X_{+-0}|^2\ \left[\matrix{|Y_{+-0}|^2 & Y_{+-0}^*\cr
                              Y_{+-0} & 1\cr}\right]\ ,\qquad
{\cal O}_{000}=|X_{000}|^2\ \left[\matrix{|Y_{000}|^2 & Y_{000}^*\cr
                              Y_{000} & 1\cr}\right]\ .
\eqno(4.4)
$$
The parameters $Y$ in (4.1) and (4.3), when expressed in terms of
the $K^0$, $\overline{K^0}$ amplitudes, are seen to depend linearly on $CP$
and $CPT$ violating terms. Therefore, in the following they will be treated
as the other small parameters of the theory; in particular, terms
containing these parameters to a degree higher than two will be dropped.

With the help of the expressions (4.2) and (4.4) one can now compute
correlations ${\cal G}$ involving pions in the final states. Keeping only
up to second order terms in all small parameters $a$, $b$, $c$, $\alpha$,
$\beta$, $\gamma$ and $Y$'s, one finds:
$$
\eqalignno{
&{\cal G}(\pi^+\pi^-,\tau_1;\pi^+\pi^-,\tau_2)={|X_{+-}|^4\over 2}\Big\{
       {\cal P}_{11}(\tau_1,\tau_2)
      +|Y_{+-}|^2\big[{\cal P}_{12}(\tau_1,\tau_2)
      +{\cal P}_{21}(\tau_1,\tau_2)\big]\cr
&\hskip 2cm
      +2|Y_{+-}|^2\, {\cal R}e\big[{\cal P}_{34}(\tau_1,\tau_2)\big]
      +2\,{\cal R}e\Big(Y_{+-}\big[{\cal P}_{13}(\tau_1,\tau_2)
      +{\cal P}_{31}(\tau_1,\tau_2)\big]\Big)\Big\}\ ,
&(4.5a)\cr
&\line{}\cr
&{\cal G}(\pi^+\pi^-,\tau_1;2\pi^0,\tau_2)=
{|X_{+-}|^2|X_{00}|^2\over 2}\Big\{
       {\cal P}_{11}(\tau_1,\tau_2)
      +|Y_{00}|^2{\cal P}_{12}(\tau_1,\tau_2)
      +|Y_{+-}|^2{\cal P}_{21}(\tau_1,\tau_2)\cr
&\hskip 2cm
      +2\, {\cal R}e\Big[
       Y_{00}\, {\cal P}_{13}(\tau_1,\tau_2)
      +Y_{+-}\, {\cal P}_{31}(\tau_1,\tau_2)
      +Y_{+-}Y^*_{00}\, {\cal P}_{34}(\tau_1,\tau_2)\Big]\Big\}\ ,
&(4.5b)\cr
&\line{}\cr
&{\cal G}(\pi^+\pi^-\pi^0,\tau_1;\pi^+\pi^-\pi^0,\tau_2)=
{|X_{+-0}|^4\over 2}\Big\{
       {\cal P}_{22}(\tau_1,\tau_2)
      +|Y_{+-0}|^2\big[{\cal P}_{12}(\tau_1,\tau_2)
      +{\cal P}_{21}(\tau_1,\tau_2)\big]\cr
&\hskip 1cm
     +2|Y_{+-0}|^2\, {\cal R}e\big[{\cal P}_{34}(\tau_1,\tau_2)\big]
      +2\, {\cal R}e\Big(Y^*_{+-0}\big[{\cal P}_{23}(\tau_1,\tau_2)
      +{\cal P}_{32}(\tau_1,\tau_2)\big]\Big)\Big\}\ ,
&(4.5c)\cr
&\line{}\cr
&{\cal G}(\pi^+\pi^-\pi^0,\tau_1;3\pi^0,\tau_2)=
{|X_{+-0}|^2|X_{000}|^2\over 2}\Big\{
       {\cal P}_{22}(\tau_1,\tau_2)
      +|Y_{+-0}|^2{\cal P}_{12}(\tau_1,\tau_2)\cr
&+|Y_{000}|^2{\cal P}_{21}(\tau_1,\tau_2)
     +2\,{\cal R}e\Big[Y^*_{+-0}\, {\cal P}_{32}(\tau_1,\tau_2)
      +Y^*_{000}\, {\cal P}_{23}(\tau_1,\tau_2)+
      Y^*_{+-0}Y_{000}\, {\cal P}_{34}(\tau_1,\tau_2)\Big]\Big\}\ .\cr
& &(4.5d)\cr
}
$$
Explicit expressions for all these quantities can be found in Appendix B.
Other correlations are 
${\cal G}(2\pi^0,\tau_1;\pi^+\pi^-,\tau_2)$ and
${\cal G}(3\pi^0,\tau_1;\pi^+\pi^-\pi^0,\tau_2)$;
they can be obtained from $(4.5b)$ and $(4.5d)$ by formally exchanging
$\tau_1$ and $\tau_2$. Further,
the probabilities ${\cal G}(2\pi^0,\tau_1;2\pi^0\tau_2)$ and
${\cal G}(3\pi^0,\tau_1;3\pi^0,\tau_2)$,
can be obtained from $(4.5a)$ and $(4.5c)$ by replacing
$X_{+-}$, $Y_{+-}$ with $X_{00}$, $Y_{00}$
and $X_{+-0}$, $Y_{+-0}$ with $X_{000}$, $Y_{000}$, respectively.

As already observed at the end of the previous section, it is of great interest
to study correlations $\cal G$ at equal proper times, in particular when the two
final states coincide. These correlations are all proportional to the
non-standard parameters $a$, $b$, $c$, $\alpha$, $\beta$, $\gamma$ and can be
used to obtain experimental estimates on these constants. 
In the case of the $\pi^+\pi^-$ final states, one finds:
$$
\eqalign{
{\cal G}(\pi^+ &\pi^-,\tau;\pi^+\pi^-,\tau)=|X_{+-}|^4\, e^{-\gamma_S\tau}
\bigg\{ e^{-\gamma_L\tau}\Big(R^L_{+-}-|\eta_{+-}|^2\Big)\cr
&-e^{-\gamma_S\tau}\bigg[{\gamma\over\Delta\Gamma}
+8\left|{C\over\Delta\Gamma_+}\right|^2-4\,{\cal R}e\bigg({\epsilon_L\,
C\over\Delta\Gamma}\bigg)\bigg]
-e^{-\Gamma\tau}\, 8\, {\cal R}e\bigg({\eta_{+-}\, C\over\Delta\Gamma_+}
e^{-i\Delta m\tau}\bigg)\bigg\}\ ,}
\eqno(4.6)
$$
where
$$
R^L_{+-}=\left|\epsilon_L+{2C^*\over\Delta\Gamma_-}+Y_{+-}\right|^2
         +{\gamma\over\Delta\Gamma}
         -8\left|{C\over\Delta\Gamma_+}\right|^2
         -4\, {\cal R}e\bigg({\epsilon_L C\over\Delta\Gamma}\bigg)\ ,\eqno(4.7a)
$$
is the $\pi^+\pi^-$ decay rate for the $K_L$ state, while
$$
\eta_{+-}=\epsilon_L-{2C^*\over\Delta\Gamma_-}+Y_{+-}\ ,\eqno(4.7b)
$$
is the parameter that measures the $K_L$, $K_S$ amplitudes ratio for the
$\pi^+\pi^-$ decay.
Similar results hold for the $\pi^+\pi^-\pi^0$-decay;
explicitly, one gets:
$$
\eqalign{
{\cal G}(\pi^+ &\pi^-\pi^0,\tau;\pi^+\pi^-\pi^0,\tau)=
|X_{+-0}|^4\, e^{-\gamma_L\tau}
\bigg\{ e^{-\gamma_S\tau}\Big(R^S_{+-0}-|\eta_{+-0}|^2\Big)\cr
&+e^{-\gamma_L\tau}\bigg[{\gamma\over\Delta\Gamma}
-8\left|{C\over\Delta\Gamma_+}\right|^2+4\,{\cal R}e\bigg({\epsilon_S\,
C^*\over\Delta\Gamma}\bigg)\bigg]
-e^{-\Gamma\tau}\, 8\, {\cal R}e\bigg({\eta_{+-0}\, C^*\over\Delta\Gamma_+}
e^{i\Delta m\tau}\bigg)\bigg\}\ ,}
\eqno(4.8)
$$
where
$$
R^S_{+-0}=\left|\epsilon_S+{2C\over\Delta\Gamma_-}+Y_{+-0}\right|^2
          -{\gamma\over\Delta\Gamma}
          -8\left|{C\over\Delta\Gamma_+}\right|^2
          -4\, {\cal R}e\bigg({\epsilon_S C^*\over\Delta\Gamma}\bigg)\eqno(4.9a)
$$
is the $CP$-violating $\pi^+\pi^-\pi^0$-decay rate of the $K_S$ state and
$$
\eta_{+-0}=\epsilon_S-{2C\over\Delta\Gamma_-}+Y_{+-0}\ ,\eqno(4.9b)
$$
measures the ratio between the $\pi^+\pi^-\pi^0$-decay amplitudes
for the $K_S$ (with isospin $I=1$) and $K_L$ states.[26] In particular, 
notice that the long time behaviour $(\tau\gg 1/\gamma_S)$ of the probability
(4.8) can give direct informations on the parameter $\gamma$:
$$
{\cal G}(\pi^+\pi^-\pi^0,\tau\,;\pi^+\pi^-\pi^0,\tau)\sim
{\gamma\over\Delta\Gamma}\ e^{-2\gamma_L\tau}\ .\eqno(4.10)
$$

Correlations among decays into semileptonic states are also of
great interest. The amplitudes for
the decay of a $K^0$ or a $\overline{K^0}$ state into $\pi^-\ell^+\nu$
and $\pi^+\ell^-\bar\nu$ are usually parametrized by three complex
constants $x$, $y$ and $z$  as follows [27]:
$$
\eqalignno{
&{\cal A}(K^0\rightarrow\pi^-\ell^+\nu)={\cal M} (1-y)\ , &(4.11a)\cr
&{\cal A}(\overline{K^0}\rightarrow\pi^+\ell^-\bar\nu)=
{\cal M}^* (1+y^*)\ , &(4.11b)\cr
&{\cal A}(K^0\rightarrow\pi^+\ell^-\bar\nu)= z\, 
{\cal A}(\overline{K^0}\rightarrow\pi^+\ell^-\bar\nu)\ , &(4.11c)\cr
&{\cal A}(\overline{K^0}\rightarrow\pi^-\ell^+\nu)=
x\, {\cal A}(K^0\rightarrow\pi^-\ell^+\nu)\ , &(4.11d) }
$$
where ${\cal M}$ is a common factor.
(In [27] $\bar x\equiv z^*$ is used instead of $z$.)
The $\Delta S=\Delta Q$ rule would forbid the decays
$K^0\rightarrow\pi^+\ell^-\bar\nu$ and 
$\overline{K^0}\rightarrow\pi^-\ell^+\nu$, so that the parameters $x$ and $z$
measure the violations of this rule. Instead, $CPT$-invariance would require
$y=\,0$.

From the above parametrization, one derive the decay amplitudes
for the states (2.1) of definite $CP$, and therefore the following
expressions for the two operators describing the semileptonic decays:[5, 19]
$$
\eqalignno{
&{\cal O}_{\ell^+}={|{\cal M}|^2\over2}\,
|1-y|^2\ \left[\matrix{|1+x|^2&(1+x^*)(1-x)\cr
&&\cr
                       (1+x)(1-x^*)&|1-x|^2\cr}\right]\ ,&(4.12a)\cr
\cr
&{\cal O}_{\ell^-}={|{\cal M}|^2\over2}\,
|1+y|^2\ \left[\matrix{|z+1|^2&(z^*+1)(z-1)\cr
&&\cr                                
                       (z+1)(z^*-1)&|z-1|^2\cr}\right]\ .&(4.12b)}
$$
The parameters $x$, $y$, and $z$ are expected to be very small, and the
available experimental determinations support this theoretical hypothesis. 
Therefore, in the following we shall treat these constants as the other small 
parameters in the theory. While the pion correlations (4.5) require 
contributions up to second order for 
comparison with the standard quantum mechanical
expressions, in the case of semileptonic decays we can limit the discussion
to the contributions up to first order in all small parameters.
Taking into account this approximation, from (3.5) and (4.12), the semileptonic
double probabilities take the form:
$$
\eqalignno{
{\cal G}(\ell^+,\tau_1;&\,\ell^-,\tau_2)={|{\cal M}|^4\over4}\bigg\{
                {\cal P}_{11}(\tau_1,\tau_2)
               +{\cal P}_{22}(\tau_1,\tau_2)
               +{\cal P}_{12}(\tau_1,\tau_2)
               +{\cal P}_{21}(\tau_1,\tau_2)\cr
&              +2\, {\cal R}e(x-z)\Big[{\cal P}_{12}(\tau_1,\tau_2)
               -{\cal P}_{21}(\tau_1,\tau_2)\Big]\cr               
&              -2\, {\cal R}e\Big[
               {\cal P}_{13}(\tau_1,\tau_2)-{\cal P}_{31}(\tau_1,\tau_2)
               +{\cal P}_{23}(\tau_1,\tau_2)-{\cal P}_{32}(\tau_1,\tau_2)\Big]
               \cr
&              -2\, {\cal R}e\Big[
               {\cal P}_{33}(\tau_1,\tau_2)
               +\big[1-2i{\cal I}m(x-z)\big]{\cal P}_{34}(\tau_1,\tau_2)
               \Big]\bigg\}
&(4.13a)\cr
{\cal G}(\ell^+,\tau_1;&\,\ell^+,\tau_2)={|{\cal M}|^4\over4}\bigg\{
                {\cal P}_{11}(\tau_1,\tau_2)
               +{\cal P}_{22}(\tau_1,\tau_2)
               +{\cal P}_{12}(\tau_1,\tau_2)
               +{\cal P}_{21}(\tau_1,\tau_2)\cr
&              -4\, {\cal R}e(y)\Big[
                 {\cal P}_{12}(\tau_1,\tau_2)
               +{\cal P}_{21}(\tau_1,\tau_2)\Big]\cr
&              +2\, {\cal R}e\Big[
                 {\cal P}_{13}(\tau_1,\tau_2)
               +{\cal P}_{31}(\tau_1,\tau_2)
               +{\cal P}_{23}(\tau_1,\tau_2)
               +{\cal P}_{32}(\tau_1,\tau_2)\Big]\cr
&              +2\, {\cal R}e\Big[
                {\cal P}_{33}(\tau_1,\tau_2)
          +\big[1-4{\cal R}e(y)\big]{\cal P}_{34}(\tau_1,\tau_2)\Big]\bigg\}
&(4.13b)\cr              
{\cal G}(\ell^-,\tau_1;&\,\ell^-,\tau_2)={|{\cal M}|^4\over4}\bigg\{
                {\cal P}_{11}(\tau_1,\tau_2)
               +{\cal P}_{22}(\tau_1,\tau_2)
               +{\cal P}_{12}(\tau_1,\tau_2)
               +{\cal P}_{21}(\tau_1,\tau_2)\cr
&              +4\, {\cal R}e(y)\Big[
                 {\cal P}_{12}(\tau_1,\tau_2)
               +{\cal P}_{21}(\tau_1,\tau_2)\Big]\cr
&              -2\, {\cal R}e\Big[
                 {\cal P}_{13}(\tau_1,\tau_2)               
               +{\cal P}_{31}(\tau_1,\tau_2)
               +{\cal P}_{23}(\tau_1,\tau_2)
               +{\cal P}_{32}(\tau_1,\tau_2)\Big]\cr
&              +2\, {\cal R}e\Big[
                 {\cal P}_{33}(\tau_1,\tau_2) 
        +\big[1+4{\cal R}e(y)\big]{\cal P}_{34}(\tau_1,\tau_2)\Big]\bigg\}\ .
&(4.13c)\cr}
$$
Explicit expressions for these correlations are collected in Appendix B.
Finally, notice that ${\cal G}(\ell^-,\tau_1;\,\ell^+,\tau_2)$ can be
formally obtained from $(4.13a)$ by exchanging $\tau_1$ and $\tau_2$.

As in the case of the pion correlations, the equal-time version of the 
above double probabilities give direct information
on the non-standard parameters $a$, $b$, $c$, $\alpha$, $\beta$ and $\gamma$.
Indeed, one finds:
$$
\eqalignno{
&{\cal G}(\ell^\pm,\tau;\ell^\pm,\tau)={|{\cal M}|^4\over4}\Bigg\{
{\gamma\over\Delta\Gamma}\big(e^{-2\gamma_L\tau}-e^{-2\gamma_S\tau}\big)\cr
&\quad
+e^{-2\Gamma \tau}\bigg[1-e^{-2(a+\alpha-\gamma)\tau}
+{a-\alpha\over\Delta m}\sin\big(2\Delta m\, \tau\big)
+{2b\over\Delta m}\Big(1-\cos(2\Delta m\,\tau)\Big)
\pm{16c\Delta\Gamma\over|\Delta\Gamma_+|^2}\bigg]\cr
&\quad
\mp8{e^{-\Gamma \tau}\over|\Delta\Gamma_+|^2}\cos\big(\Delta m\,\tau\big)\bigg[
 c\Delta\Gamma\big(e^{-\gamma_S\tau}+e^{-\gamma_L\tau}\big)
+2\beta\Delta m\big(e^{-\gamma_S\tau}-e^{-\gamma_L\tau}\big)\bigg]\cr
&\quad
\mp8{e^{-\Gamma \tau}\over|\Delta\Gamma_+|^2}\sin\big(\Delta m\,\tau\big)\bigg[
 \beta\Delta\Gamma\big(e^{-\gamma_S\tau}+e^{-\gamma_L\tau}\big)
-2c\Delta m\big(e^{-\gamma_S\tau}-e^{-\gamma_L\tau}\big)\bigg]\Bigg\}
&(4.14)\cr
\cr
&{\cal G}(\ell^\pm,\tau;\ell^\mp,\tau)={|{\cal M}|^4\over4}\Bigg\{
{\gamma\over\Delta\Gamma}\big(e^{-2\gamma_L\tau}-e^{-2\gamma_S\tau}\big)\cr
&\quad
+e^{-2\Gamma \tau}\bigg[1+e^{-2(a+\alpha-\gamma)\tau}
-{a-\alpha\over\Delta m}\sin\big(2\Delta m\,\tau\big)
-{2b\over\Delta m}\Big(1-\cos\big(2\Delta m\,\tau\big)\Big)\bigg]\Bigg\}
&(4.15)\cr
}
$$
In particular, the small time behaviour of these probabilities
allow to obtain estimates on the non-standard parameter $a>0$:
$$
{{\cal G}(\ell^\pm,\tau;\ell^\pm,\tau)\over
{\cal G}(\ell^\pm,\tau;\ell^\mp,\tau)}\sim
2\ a\ \tau\ .\eqno(4.16)
$$

All the correlation probabilities ${\cal G}(f_1,\tau_1;f_2,\tau_2)$ 
discussed so far can be measured
at a $\phi$-factory, and bounds on the various non-standard parameters
$a$, $b$, $c$, $\alpha$, $\beta$, $\gamma$ can be obtained, at least
in principle, by fitting the different time-behaviours. However, much of the
analysis of $\phi$-decay experiments has been carried out using integrated
distributions at fixed time interval $\tau=\tau_1-\tau_2$.[21-24, 28]
In the generic case, these single-time probabilities are defined as:
$$
{\cal F}(f_1,f_2;\tau)=\int_0^\infty dt\, {\cal G}(f_1,t+\tau;f_2,t)\ ,
\eqno(4.17)
$$
where $\tau$ is taken to be positive. For negative $\tau$, one defines:
$$
{\cal F}(f_1,f_2;-|\tau|)=\int_0^\infty dt\, {\cal G}(f_1,t-|\tau|;f_2,t)\ 
\theta(t-|\tau|)\ ;
\eqno(4.18)
$$
the presence of the step-function is necessary since the evolution is
of semigroup type, with forward in time propagation, starting at the origin
(we can not propagate a kaon before it is created in a $\phi$-decay).
In this case, one easily finds: ${\cal F}(f_1,f_2;-|\tau|)=
{\cal F}(f_2,f_1;|\tau|)$. In the following, we shall always assume:
$\tau\geq0$.

With the integrated probabilities $\cal F$ one can form asymmetries that are
sensitive to various parameters in the theory. For example, the following
asymmetry involving two-pion decays:[28]
$$
{\cal A}_{\epsilon^\prime}(\tau)={
{\cal F}(\pi^+\pi^-,2\pi^0;\tau) - {\cal F}(2\pi^0,\pi^+\pi^-;\tau)\over
{\cal F}(\pi^+\pi^-,2\pi^0;\tau) + {\cal F}(2\pi^0,\pi^+\pi^-;\tau) }\ ,
\eqno(4.19)
$$
is particularly useful in the determination of the ratio
$\epsilon^\prime/\epsilon$, where $\epsilon$ and $\epsilon^\prime$ are
the phenomenological constants that parametrize the $K_L$, $K_S$
two-pion decay amplitude ratios (see $(4.7b)$ and the analogous expression
in the case of neutral pions):
$$
\eta_{+-}=\epsilon+\epsilon^\prime\ ,\qquad\quad
\eta_{00}=\epsilon-2\, \epsilon^\prime\ .\eqno(4.20)
$$
Using the relations in Appendix B, one can show that, to first order in
$\epsilon^\prime/\epsilon$:
$$
{\cal A}_{\epsilon^\prime}(\tau)=3\, 
{\cal R}e\Big({\epsilon^\prime\over\epsilon}\Big)\ {N_1(\tau)\over D(\tau)}
-3\, {\cal I}m\Big({\epsilon^\prime\over\epsilon}\Big)\ 
{N_2(\tau)\over D(\tau)}\ .\eqno(4.21)
$$
The $\tau$-dependent coefficients $N_1$, $N_2$ and $D$ are functions of the
non-standard parameters $c$, $\beta$, $\gamma$ and of $\eta_{+-}$;
their explicit expressions can be found in Appendix C.

From the experimental determination of the $\tau$-dependence of
${\cal A}_{\epsilon^\prime}(\tau)$ one can extract, at least in principle,
the real and imaginary part of $\epsilon^\prime/\epsilon$;
however, notice that this is much more problematic than in the standard
quantum mechanical case, in which $N_1$, $N_2$ and $D$ have very simple
$\tau$-dependences. This has already been observed in [5] for the case
of a simply-positive kaon dynamics,  and remains true in the large
time limit $(\tau\gg 1/\gamma_S)$; in this limit, (4.21) becomes:
$$
\eqalign{
{\cal A}_{\epsilon^\prime}\sim 
3\, {\cal R}e\Big({\epsilon^\prime\over\epsilon}\Big)\ 
\bigg\{1&-{1\over|\eta_{+-}|^2}\bigg[ {\gamma\over\Delta\Gamma}
-8\left|{C\over\Delta\Gamma_+}\right|^2
-4\, {\cal R}e\bigg({\epsilon_L C\over\Delta\Gamma}\bigg)\bigg]\cr
&-4\, {\cal R}e\bigg({C\over\Delta\Gamma_+\, \eta_{+-}^*}\bigg)\bigg\}
-12\, {\cal I}m\Big({\epsilon^\prime\over\epsilon}\Big)\
{\cal I}m\bigg({C\over\Delta\Gamma_+\, \eta_{+-}^*}\bigg)\ .}
\eqno(4.22)
$$

Similar results hold in the case of asymmetries involving semileptonic decays.
The following three cases are of particular interest:[5, 28]
$$
\eqalignno{
&{\cal A}_L(\tau)={
{\cal F}(\ell^+,\ell^+;\tau) - {\cal F}(\ell^-,\ell^+;\tau)\over
{\cal F}(\ell^+,\ell^+;\tau) + {\cal F}(\ell^-,\ell^-;\tau) }\ , &(4.23a)\cr
&{\cal A}_T(\tau)={
{\cal F}(\ell^+,\ell^+;\tau) - {\cal F}(\ell^-,\ell^-;\tau)\over
{\cal F}(\ell^+,\ell^+;\tau) + {\cal F}(\ell^-,\ell^-;\tau) }\ , &(4.23b)\cr
&{\cal A}_{CPT}(\tau)={
{\cal F}(\ell^+,\ell^-;\tau) - {\cal F}(\ell^-,\ell^+;\tau)\over
{\cal F}(\ell^+,\ell^+;\tau) + {\cal F}(\ell^-,\ell^-;\tau) }\ . &(4.23c)\cr}
$$
The first asymmetry is connected with the so called $CP$-violating charge
asymmetry (see [28]), while ${\cal A}_T$ and 
${\cal A}_{CPT}$ signal direct violation of time-reversal and $CPT$ invariance.
For simplicity, we again collect here only the large-time limit
$(\tau\gg 1/\gamma_S)$ of their expressions:
$$
\eqalignno{
&{\cal A}_L=2\, {\cal R}e\bigg(\epsilon_L +{2 C^*\over\Delta\Gamma_-}\bigg)
-{\cal R}e\big(x-z+2\, y\big)\ , &(4.24a)\cr
&{\cal A}_T=2\, {\cal R}e\bigg(\epsilon_L +{2 C^*\over\Delta\Gamma_-}\bigg)
+2\, {\cal R}e\bigg(\epsilon_S +{2 C\over\Delta\Gamma_-}\bigg)\cr
&\hskip 4.5cm -4\, {\cal R}e(y)
-16\,{\cal R}e\bigg({\Gamma\, C\over\Delta\Gamma_-(\Gamma_+ +\gamma_L)}\bigg)
\ ,&(4.24b)\cr
&{\cal A}_{CPT}=2\, {\cal R}e\bigg(\epsilon_L 
+{2 C^*\over\Delta\Gamma_-}\bigg)
-2\, {\cal R}e\bigg(\epsilon_S +{2 C\over\Delta\Gamma_-}\bigg)\cr
&\hskip 4.5cm -2\, {\cal R}e(x-z)
+16\,{\cal R}e\bigg({\Gamma\, C\over\Delta\Gamma_-(\Gamma_+ +\gamma_L)}\bigg)
\ .&(4.24c)}
$$
Notice that the asymptotic value for ${\cal A}_L$ coincides with the
charge asymmetry $\delta_L$ for the $K_L$ state (see [19]).

Of great interest are also mixed asymmetries, involving both pion and
semileptonic decays. These are particularly useful for studying $CP$-violating
effects in $K_S\rightarrow 3\pi$ decays, which are hard to detect directly,
due to the smallness of their branching ratios. 
Indeed, as suggested in [29, 28],
one can alternatively study the following asymmetry, integrated over
the whole Dalitz plot:
$$
{\cal A}_{3\pi}(\tau)={
{\cal F}(3\pi,\ell^+;\tau) - {\cal F}(3\pi,\ell^-;\tau)\over
{\cal F}(3\pi,\ell^+;\tau) + {\cal F}(3\pi,\ell^-;\tau) \ .}
\eqno(4.25)
$$
In the case of the $3\pi^0$ decay, one explicitly obtains:
$$
{\cal A}_{3\pi^0}(\tau)={N_{3\pi^0}(\tau)\over D_{3\pi^0}(\tau)}\ ,
\eqno(4.26)
$$
where
$$
\eqalignno{
&N_{3\pi^0}(\tau)=2\, {\cal R}e\bigg(\epsilon_S +{2 C\over\Delta\Gamma_-}
-{8\Gamma\, C\over\Delta\Gamma_-(\Gamma_+ +\gamma_L)}\bigg)
+{\cal R}e\big(x-z-2\,y\big)\hskip 3cm\cr
&\hskip .5cm -e^{-\Delta\Gamma\tau}\bigg({\gamma\over\Delta\Gamma}\bigg)\bigg[
2\, {\cal R}e\bigg(\epsilon_L +{2 C\over\Delta\Gamma_+}
-{8\Gamma\, C\over\Delta\Gamma_+(\Gamma_+ +\gamma_S)}\bigg)
-{\cal R}e\big(x-z+2\,y\big)\bigg] &(4.27a)\cr
&\hskip .5cm -2\, e^{-\Delta\Gamma\tau/2}\,
{\cal R}e\Big(\eta_{000}\, e^{i\Delta m\tau}\Big)
\ ,\cr
&\cr
&D_{3\pi^0}(\tau)=1+ e^{-\Delta\Gamma\tau}\bigg[ R^S_{000}+
{\gamma\over\Delta\Gamma}\bigg({\cal R}e\big(x+z\big)
+{\gamma\over\Delta\Gamma}\, {\gamma_L\over\gamma_S}\bigg)\bigg]\ , &(4.27b)}
$$
and $R^S_{000}$ and $\eta_{000}$ are obtained from $(4.9a)$ and $(4.9b)$
by replacing the parameter $Y_{+-0}$ with $Y_{000}$. A similar expression holds
for the $\pi^+\pi^-\pi^0$ asymmetry, provided one performs an even integration
over the pion Dalitz plot. These asymmetries should be accessible
to the experiment at a $\phi$-factory.

\vskip 1cm

{\bf 5. DISCUSSION}
\medskip

As discussed in the previous section, the dynamics of the neutral kaon system 
based on the non-standard quantum evolution equation (2.3) can be probed at
a $\phi$-factory by measuring the time-evolution of various observables.
Although the most striking differences with respect to the predictions of
ordinary quantum mechanics arise at equal proper times, all the double
probabilities computed in Section 4 have a non-trivial dependence
on the parameters $a$, $b$, $c$, $\alpha$, $\beta$ and $\gamma$,
with a corresponding characteristic time-behaviour. 
Therefore, by fitting the analytic expressions of these observables
with the experimental data, it will be possible to obtain estimates
for these parameters and, as a consequence, check the inequalities (2.11)
they have to satisfy. Indeed, on phenomenological grounds,
the values of the non-standard parameters 
may be expected to be of the order $m_k^2/m_P\sim 10^{-19} {\rm GeV}$,
with $m_k$ the kaon mass and 
$m_P$ the Planck mass, and therefore within the reach of an high
luminosity $\phi$-factory.[5]
It should be stressed that having bounds on the values of
$a$, $b$, $c$, $\alpha$, $\beta$ and $\gamma$ is important also for the
measure of the other parameters of the model, {\it e.g.}
the ratio $\epsilon^\prime/\epsilon$. In fact, the determination of the
real and the imaginary part of this ratio via (4.21) and (4.22) is
meaningful provided estimates on the non-standard parameters
have been independently given.

A complete analysis on the sensitivities in the measure of
$a$, $b$, $c$, $\alpha$, $\beta$ and $\gamma$ that can be reached in an
actual experimental setup at a $\phi$-factory requires careful estimates
of various effects, like the finite size of the detectors, and are
certainly beyond the scope of the present paper. Nevertheless, we
would like to comment on a possible background effect that could in principle
interfere with the determination of $a$, $b$, $c$, $\alpha$, $\beta$ and
$\gamma$.

Although not yet observed, the radiative decay 
$\phi\rightarrow \gamma\, K^0 \overline{K^0}$ could constitute a non-vanishing
source of background noise in a $\phi$-factory experiment.[28] 
Due to the presence
of the photon in the final state, the system of the two correlated neutral 
kaons are no longer initially described by (3.1), but rather by the symmetric
state:
$$
|\psi_S\rangle= {1\over\sqrt2}\Big(|K_1,-p\rangle \otimes  |K_2,p\rangle +
|K_2,-p\rangle \otimes  |K_1,p\rangle\Big)\ ,\eqno(5.1)
$$
or equivalently by the density matrix:
$$
\rho_S={1\over 2}\Big[P_1\otimes P_1\ +\ P_2\otimes P_2\ -\ 
P_3\otimes P_3\ -\ P_4\otimes P_4\Big]\ .\eqno(5.2)
$$

Once created in a $\phi$-decay, the state $\rho_S$ will also evolve according
with the completely positive dynamics regulated by the equation (2.3)
and contribute to the double probability distributions
${\cal G}(f_1,\tau_1;f_2,\tau_2)$ considered in the previous section.
Although these additional contributions are suppressed by the branching ratio
$$
r={{\cal A}(\phi\rightarrow\gamma K^0 \overline{K^0})\over
{\cal A}(\phi\rightarrow K^0 \overline{K^0})}\ ,\eqno(5.3)
$$
they could give rise to non-negligible effects. 

In order to check this possibility, we have computed the $\rho_S$-contributions
to the three probabilities distributions $\cal G$ involving equal final states
($\pi^+\pi^-$, $\pi^+\pi^-\pi^0$ and $\pi^-\ell^+\nu$); 
the explicit expressions are collected
in Appendix D. One observes that the dependence on the parameters of the model,
as well as the time-dependence of these contributions differ from the ones
described in the previous section. Therefore, at least in principle, it is
possible to isolate the background $\rho_S$-contribution from the interesting
$\rho_A$-terms.

However, if the branching ratio $r$ is of order $10^{-7}-10^{-8}$, 
as some theoretical investigations seem to suggest,[30-32] 
there are instances where the symmetric
background contribution can not be neglected. This is the case for the
asymptotic long-time behaviour of the equal-time probability
${\cal G}(\pi^+\pi^-\pi^0,\tau\,;\pi^+\pi^-\pi^0,\tau)$; using the results of
Appendix D, in the presence of the $\rho_S$-background the asymptotic behaviour
in (4.10) is modified as follows: 
$$
{\cal G}(\pi^+\pi^-\pi^0,\tau\,;\pi^+\pi^-\pi^0,\tau)\sim
\left({\gamma\over\Delta\Gamma}+{r\over2}\right)
\ e^{-2\gamma_L\tau}\ ,\eqno(5.4)
$$
and one needs to study the full double-time correlation
${\cal G}(\pi^+\pi^-\pi^0,\tau_1\,;\pi^+\pi^-\pi^0,\tau_2)$ in order
to get an estimate for the parameter $\gamma$. On the other hand,
the correction to the ratio of semileptonic probabilities in (4.16)
is proportional to $r$ times terms linear in the non-standard parameters,
and therefore can be totally ignored. One should also notice that the effect
of the background $\phi\rightarrow \gamma K^0 \overline{K^0}$ can be
greatly reduced by a careful choice of the geometry of 
the measuring detector.[33]
In conclusion, a high-luminosity $\phi$-factory can put stringent limits
on the values of the parameters 
$a$, $b$, $c$, $\alpha$, $\beta$ and $\gamma$, while keeping under control
possible background effects.

As a final remark, we would like to point out another unique feature
of the $\phi$-interferometry physics. All the considerations of the
previous sections are based on the assumption of a completely positive
dynamics for the neutral kaon system. Had we used a simply positive
time-evolution instead, inconsistencies in the formalism would have emerged
in the study of correlated kaons. One of the properties of the density
matrix $\rho_A(\tau)\equiv\rho_A(\tau,\tau)$ in (3.4) is that its
mean value on any vector is positive for all times. Without this basic
requirement, the standard probability interpretation of $\rho_A(\tau)$
as a state of the correlated kaons would be meaningless.

In the case of the vector:
$$
|u\rangle={1\over\sqrt2}\left[
\left({1\atop0}\right) \otimes \left({1\atop0}\right)+
\left({0\atop1}\right) \otimes \left({0\atop1}\right)\right]\ ,\eqno(5.5)
$$
the completely positive dynamics of equation (2.3) gives, to first order
in the small parameters:
$$
\eqalign{
&{\cal U}(\tau)\equiv\langle u|\rho_A(\tau)|u\rangle=\bigg\{
{\gamma\over{2\Delta\Gamma}}
\left(e^{-2\gamma_L\tau}\ -\ e^{-2\gamma_S\tau}\right)\cr
&\hskip 1cm
-e^{-2\Gamma\tau}\ \left[
{{\alpha-a}\over{2\Delta m}}\ \sin\big(2\Delta m\, \tau\big)\ +\
{b\over{\Delta m}}\Big(\cos\big(2\Delta m\, \tau\big)-1\Big)\right]\bigg\}\ .}
\eqno(5.6)
$$
This result is indeed positive for all times, and vanish only at $\tau=\,0$,
due to the antisymmetry of $\rho_A$ in (3.3). 

This conclusion is not true in general if the time evolution for $\rho_A$
is only positive and not completely positive. For instance, if one chooses
to work with the simply positive dynamics studied in [1-4, 5], that can be
obtained from the one in (2.3), (2.10) by setting $a=b=c=\,0$, 
then the mean value ${\cal U}(\tau)$ would be negative
for small enough times, signaling the presence of unphysical negative
probabilities (see also [19]).
A $\phi$-factory, being a high performance quantum interferometer, can measure,
at least in principle, observables like (5.6) and therefore 
further clarify the request of complete positivity for the dynamics
of neutral kaons.

\vfill\eject

{\bf APPENDIX A}
\medskip

In order to calculate the explicit expressions of the observables 
${\cal G}(f_1,\tau_1;f_2,\tau_2)$ in (3.5)
we need to determine the time evolution of the single kaon operators 
$P_i$, $i=1,2,3,4$, in (3.2). Using (2.13), with (3.2) as initial conditions,
one finds 
$$
\eqalignno{
&P_1(\tau)=\left(
\matrix{M^{(0)}_{11}(\tau)+M^{(2)}_{11}(\tau)
&       M^{(1)}_{31}(\tau)+M^{(2)}_{31}(\tau)\cr\cr
        M^{(1)}_{41}(\tau)+M^{(2)}_{41}(\tau)
&       M^{(1)}_{21}(\tau)+M^{(2)}_{21}(\tau)}
\right)
&(A.1a)\cr
&\cr
&\cr
&P_2(\tau)=\left(
\matrix{M^{(1)}_{12}(\tau)+M^{(2)}_{12}(\tau)
&       M^{(1)}_{32}(\tau)+M^{(2)}_{32}(\tau)\cr\cr
       M^{(1)}_{42}(\tau)+M^{(2)}_{42}(\tau)
&       M^{(0)}_{22}(\tau)+M^{(2)}_{22}(\tau)}
\right)
&(A.1b)\cr
&\cr
&\cr
&P_3(\tau)=\left(
\matrix{M^{(1)}_{13}(\tau)+M^{(2)}_{13}(\tau)
&       M^{(0)}_{33}(\tau)+M^{(2)}_{33}(\tau)\cr\cr
      M^{(1)}_{43}(\tau)+M^{(2)}_{43}(\tau)
&       M^{(1)}_{23}(\tau)+M^{(2)}_{23}(\tau)}
\right)
&(A.1c)\cr
&\cr
&\cr
&P_4(\tau)=\left(
\matrix{M^{(1)}_{14}(\tau)+M^{(2)}_{14}(\tau)
&       M^{(1)}_{34}(\tau)+M^{(2)}_{34}(\tau)\cr\cr
      M^{(0)}_{44}(\tau)+M^{(2)}_{44}(\tau)
&       M^{(1)}_{24}(\tau)+M^{(2)}_{24}(\tau)}
\right)\ .&(A.1d)}
$$
With these expressions, one can now calculate the ``elementary'' double
probabilities in (3.8):
$$
\eqalign{
{\cal P}_{ij}(\tau_1,\tau_2)
&=\hbox{Tr}\big\{P_i\, P_1(\tau_1)\big\}\
\hbox{Tr}\big\{P_j\, P_2(\tau_2)\big\}\ +\
\hbox{Tr}\big\{P_i\, P_2(\tau_1)\big\}\ 
\hbox{Tr}\big\{P_j\, P_1(\tau_2)\big\}\cr
&\qquad
-\hbox{Tr}\big\{P_i\, P_3(\tau_1)\big\}\
\hbox{Tr}\big\{P_j\, P_4(\tau_2)\big\}\ -\
\hbox{Tr}\big\{P_i\, P_4(\tau_1)\big\}\
\hbox{Tr}\big\{P_j\, P_3(\tau_2)\big\}\ .
}\eqno(A.2)
$$
They satisfy the following symmetry property:
$$
{\cal P}_{ij}(\tau_1,\tau_2)={\cal P}_{ji}(\tau_2,\tau_1)\ ;\eqno(A.3)
$$
further, since the single kaon dynamics preserves hermiticity,
$$
\eqalignno{
&{\cal P}_{ij}(\tau_1,\tau_2)={\cal P}_{ij}(\tau_1,\tau_2)^*\ ,
\quad i,j=1,2\ , &(A.4a)\cr
&{\cal P}_{ij}(\tau_1,\tau_2)^*={\cal P}_{ji}(\tau_1,\tau_2)\ ,
\quad i,j=3,4\ , &(A.4b)\cr
&{\cal P}_{i3}(\tau_1,\tau_2)^*={\cal P}_{i4}(\tau_1,\tau_2)\ ,
\quad i=1,2\ . &(A.4c)}
$$
\vfill\eject
It then follows:
$$
\eqalignno{
&{\cal P}_{11}(\tau_1,\tau_2)=
                    M^{(0)}_{11}(\tau_1)\ M^{(1)}_{12}(\tau_2)
                   +M^{(1)}_{12}(\tau_1)\ M^{(0)}_{11}(\tau_2)\cr
&\phantom{{\cal P}_{11}(\tau_1,\tau_2)}
                   +M^{(0)}_{11}(\tau_1)\ M^{(2)}_{12}(\tau_2)
                   +M^{(2)}_{12}(\tau_1)\ M^{(0)}_{11}(\tau_2) &(A.5a)\cr
&\phantom{{\cal P}_{11}(\tau_1,\tau_2)}
                   -2\,{\cal R}e\Big(M^{(1)}_{13}(\tau_1)\ 
                   M^{(1)}_{14}(\tau_2)\Big)
\cr
&{\cal P}_{12}(\tau_1,\tau_2)=
                    M^{(0)}_{11}(\tau_1)\ M^{(0)}_{22}(\tau_2)\cr 
&\phantom{{\cal P}_{11}(\tau_1,\tau_2)}
                   +M^{(0)}_{11}(\tau_1)\ M^{(2)}_{22}(\tau_2)
                   +M^{(2)}_{11}(\tau_1)\ M^{(0)}_{22}(\tau_2) &(A.5b)\cr
&\phantom{{\cal P}_{11}(\tau_1,\tau_2)}
                   +M^{(1)}_{12}(\tau_1)\ M^{(1)}_{21}(\tau_2)
                   -2\, {\cal R}e\ \Big(M^{(1)}_{13}(\tau_1)\ 
                   M^{(1)}_{24}(\tau_2)\Big)
\cr
&{\cal P}_{13}(\tau_1,\tau_2)=
                    M^{(0)}_{11}(\tau_1)\ M^{(1)}_{42}(\tau_2)
                   -M^{(1)}_{13}(\tau_1)\ M^{(0)}_{44}(\tau_2)\cr 
&\phantom{{\cal P}_{11}(\tau_1,\tau_2)}
                   +M^{(0)}_{11}(\tau_1)\ M^{(2)}_{42}(\tau_2)
                   -M^{(2)}_{13}(\tau_1)\ M^{(0)}_{44}(\tau_2) &(A.5c)\cr
&\phantom{{\cal P}_{11}(\tau_1,\tau_2)}
                   +M^{(1)}_{12}(\tau_1)\ M^{(1)}_{41}(\tau_2)
                   -M^{(1)}_{14}(\tau_1)\ M^{(1)}_{43}(\tau_2)
\cr
&{\cal P}_{14}(\tau_1,\tau_2)={\cal P}_{13}(\tau_1,\tau_2)^*
&(A.5d)\cr
\cr
&{\cal P}_{21}(\tau_1,\tau_2)={\cal P}_{12}(\tau_2,\tau_1)
&(A.6a)\cr
&{\cal P}_{22}(\tau_1,\tau_2)= 
                    M^{(1)}_{21}(\tau_1)\ M^{(0)}_{22}(\tau_2)
                   +M^{(0)}_{22}(\tau_1)\ M^{(1)}_{21}(\tau_2)\cr
&\phantom{{\cal P}_{11}(\tau_1,\tau_2)}
                   +M^{(2)}_{21}(\tau_1)\ M^{(0)}_{22}(\tau_2)
                   +M^{(0)}_{22}(\tau_1)\ M^{(2)}_{21}(\tau_2) &(A.6b)\cr
&\phantom{{\cal P}_{11}(\tau_1,\tau_2)}
                   -2\,{\cal R}e\ \Big(M^{(1)}_{23}(\tau_1)\ 
                   M^{(1)}_{24}(\tau_2)\Big)
\cr
&{\cal P}_{23}(\tau_1,\tau_2)=
                    M^{(0)}_{22}(\tau_1)\ M^{(1)}_{41}(\tau_2)
                   -M^{(1)}_{23}(\tau_1)\ M^{(0)}_{44}(\tau_2)\cr 
&\phantom{{\cal P}_{11}(\tau_1,\tau_2)}
                   +M^{(0)}_{22}(\tau_1)\ M^{(2)}_{41}(\tau_2)
                   -M^{(2)}_{23}(\tau_1)\ M^{(0)}_{44}(\tau_2) &(A.6c)\cr
&\phantom{{\cal P}_{11}(\tau_1,\tau_2)}
                   +M^{(1)}_{21}(\tau_1)\ M^{(1)}_{42}(\tau_2)
                   -M^{(1)}_{24}(\tau_1)\ M^{(1)}_{43}(\tau_2)
\cr
&{\cal P}_{24}(\tau_1,\tau_2)={\cal P}_{23}(\tau_1,\tau_2)^*
&(A.6d)\cr
\cr
&{\cal P}_{31}(\tau_1,\tau_2)={\cal P}_{13}(\tau_2,\tau_1)
&(A.7a)\cr
&{\cal P}_{32}(\tau_1,\tau_2)={\cal P}_{23}(\tau_2,\tau_1)
&(A.7b)\cr
&{\cal P}_{33}(\tau_1,\tau_2)=
                   -M^{(1)}_{43}(\tau_1)\ M^{(0)}_{44}(\tau_2)
                   -M^{(0)}_{44}(\tau_1)\ M^{(1)}_{43}(\tau_2)\cr
&\phantom{{\cal P}_{11}(\tau_1,\tau_2)}
                   -M^{(2)}_{43}(\tau_1)\ M^{(0)}_{44}(\tau_2) 
                   -M^{(0)}_{44}(\tau_1)\ M^{(2)}_{43}(\tau_2) &(A.7c)\cr
&\phantom{{\cal P}_{11}(\tau_1,\tau_2)}
                   +M^{(1)}_{41}(\tau_1)\ M^{(1)}_{42}(\tau_2)
                   +M^{(1)}_{42}(\tau_1)\ M^{(1)}_{41}(\tau_2)
\cr
&{\cal P}_{34}(\tau_1,\tau_2)=
                   -M^{(0)}_{44}(\tau_1)\ M^{(0)}_{33}(\tau_2)\cr
&\phantom{{\cal P}_{11}(\tau_1,\tau_2)}
                   -M^{(0)}_{44}(\tau_1)\ M^{(2)}_{33}(\tau_2)
                   -M^{(2)}_{44}(\tau_1)\ M^{(0)}_{33}(\tau_2) &(A.7d)\cr
&\phantom{{\cal P}_{11}(\tau_1,\tau_2)}
                   +M^{(1)}_{41}(\tau_1)\ M^{(1)}_{32}(\tau_2) 
                   +M^{(1)}_{42}(\tau_1)\ M^{(1)}_{31}(\tau_2)
                   -M^{(1)}_{43}(\tau_1)\ M^{(1)}_{34}(\tau_2)
\cr
\cr
&{\cal P}_{41}(\tau_1,\tau_2)={\cal P}_{31}(\tau_1,\tau_2)^*
&(A.8a)\cr
&{\cal P}_{42}(\tau_1,\tau_2)={\cal P}_{32}(\tau_1,\tau_2)^*
&(A.8b)\cr
&{\cal P}_{43}(\tau_1,\tau_2)={\cal P}_{34}(\tau_2,\tau_1)
&(A.8c)\cr
&{\cal P}_{44}(\tau_1,\tau_2)={\cal P}_{33}(\tau_1,\tau_2)^*
\ .
&(A.8d)\cr
}
$$
Using the expressions for $M_{ij}^{(0)}$ and $M_{ij}^{(1)}$ given in Section 2,
together with the ones collected in the Appendix of [17] for $M_{ij}^{(2)}$,
one arrives at the formulas listed below. As for the entries of
$M_{ij}^{(0)}$, $M_{ij}^{(1)}$ and $M_{ij}^{(2)}$, in order to preserve
the exponential form of the time evolution, linear and quadratic corrections in
$\tau_1$ and $\tau_2$ have been reabsorbed in the exponents.
$$
\eqalignno{
&{\cal P}_{11}(\tau_1,\tau_2)=
        e^{-\gamma_S(\tau_1+\tau_2)}\bigg[
       -2{\gamma\over\Delta\Gamma} 
       +{8\over\Delta\Gamma}{\cal R}e(C\epsilon_L)
       -16\left|{C\over\Delta\Gamma_+}\right|^2\bigg]\cr
&\phantom{{\cal P}_{11}}
       +\bigg(e^{-\gamma_S \tau_1-\gamma_L \tau_2}
       +e^{-\gamma_L \tau_1-\gamma_S \tau_2}\bigg)
       \bigg[{\gamma\over\Delta\Gamma}
       +|e_L(+,-)|^2
       -{4\over\Delta\Gamma}{\cal R}e(C\epsilon_L)
       -8\left|{C\over\Delta\Gamma_+}\right|^2\bigg]\cr
&\phantom{{\cal P}_{11}}
       -{\cal R}e\bigg[\bigg(e^{-\gamma_S \tau_1-\Gamma_+ \tau_2}
       +e^{-\Gamma_+ \tau_1-\gamma_S \tau_2}\bigg)\
        {8C\over\Delta\Gamma_+}e_L(-,-)\bigg]\cr
&\phantom{{\cal P}_{11}}
       -2\, {\cal R}e\bigg[e^{-\Gamma_+\tau_1-\Gamma_-\tau_2}\ 
       |e_L(-,-)|^2\bigg]
&(A.9a)\cr
\cr
&{\cal P}_{12}(\tau_1,\tau_2)=\bigg(
      2\, e^{-\gamma_S(\tau_1+\tau_2)}
         +e^{-\gamma_L(\tau_1+\tau_2)}
       -e^{-\gamma_L \tau_1-\gamma_S \tau_2}\bigg) 
        \bigg({\gamma\over\Delta\Gamma}\bigg)^2\cr
&\phantom{{\cal P}_{12}}
       +e^{-\gamma_S \tau_1-\gamma_L \tau_2}\bigg[
        1-3\bigg({\gamma\over\Delta\Gamma}\bigg)^2
        +2\, {\cal R}e\bigg(\epsilon_S\epsilon_L
       +2{C^*\epsilon_S+C\epsilon_L\over\Delta\Gamma_-}\bigg)
       -24\, {\cal R}e\bigg({|C|^2\over(\Delta\Gamma_+)^2}\bigg)\bigg]\cr
&\phantom{{\cal P}_{12}}
       -{\cal R}e\bigg[e^{-\gamma_S \tau_1-\Gamma_- \tau_2}\
        {8C^*\over\Delta\Gamma_-}f_S(-,-)\bigg]
       -{\cal R}e\bigg[e^{-\Gamma_- \tau_1-\gamma_L \tau_2}\
        {8C^*\over\Delta\Gamma_+}e^*_L(-,-)\Big)\bigg]\cr
&\phantom{{\cal P}_{12}}
    -2\, {\cal R}e\bigg[e^{-\Gamma_+\tau_1-\Gamma_-\tau_2}\ 
    e_L(-,-)f_S(-,-)\bigg]
&(A.9b)\cr
\cr
&{\cal P}_{13}(\tau_1,\tau_2)=\bigg(e^{-\gamma_L\tau_1-\gamma_S\tau_2}
       -2\,e^{-\gamma_S(\tau_1+\tau_2)}\bigg)\ 
       {\gamma\over\Delta\Gamma}\, f_S(+,-) \cr
&\phantom{{\cal P}_{12}}
       +e^{-\gamma_S \tau_1-\gamma_L \tau_2}\bigg[
       \bigg(1-{2(A-\gamma)\over\Delta\Gamma_+}\bigg)e^*_L(+,-)
       +{2B^*\over\Delta\Gamma_+}e_L(+,-)
       -{\gamma\over\Delta\Gamma}{\Delta\Gamma_-\over\Delta\Gamma_+}f_S(+,-)
       \bigg]\cr
&\phantom{{\cal P}_{12}}
       -e^{-\gamma_S \tau_1-(\Gamma_+ +A-\gamma)\tau_2}\bigg[
       4{C\over\Delta\Gamma_+}
       -{4(A-\gamma)\over\Delta\Gamma_+}\epsilon_L^*
       -{4\gamma\epsilon_S\over\Delta\Gamma_+}
       -{4\gamma C\over\Delta\Gamma\Delta\Gamma_-}
       +{4B^*\epsilon_L\over\Delta \Gamma_+}
       +{2iB^*C^*\over\Delta m\Delta\Gamma_-}\bigg]\cr
&\phantom{{\cal P}_{12}}
       +e^{-\gamma_S\tau_1-\Gamma_- \tau_2}\
          {2iB^*C^*\over\Delta m\Delta\Gamma_-}
       -e^{-\gamma_L\tau_1-\Gamma_+\tau_2}\
          {4\gamma C\over\Delta\Gamma\Delta\Gamma_-}\cr
&\phantom{{\cal P}_{12}}
       -e^{-\Gamma_-\tau_1-\Gamma_+\tau_2-(A-\gamma)(\tau_1+\tau_2)}\bigg[
       \bigg(1+{2(A-\gamma)\over\Delta\Gamma_+}\bigg)e^*_L(-,-)
       +{2\gamma\over\Delta\Gamma_+}f_S(-,-)\cr
&\phantom{{\cal P}_{12}}
-{iB^*\over2\Delta m}{\Delta\Gamma_-\over\Delta\Gamma_+}e_L(-,-)\bigg]
       +e^{-\Gamma_+\tau_1-\Gamma_-\tau_2}\ {iB^*\over2\Delta m}e_L(-,-)
       -e^{-\Gamma_+(\tau_1+\tau_2)}\ {iB^*\over\Delta m}e_L(-,-)\cr
&&(A.9c)\cr
\cr
&{\cal P}_{22}(\tau_1,\tau_2)=\bigg(e^{-\gamma_S \tau_1-\gamma_L \tau_2}
        +e^{-\gamma_L \tau_1-\gamma_S \tau_2}\bigg)\bigg[
       |f_S(+,-)|^2-{\gamma\over\Delta\Gamma}
       -{4\over\Delta\Gamma}{\cal R}e(C^*\epsilon_S)
       -8\left|{C\over\Delta\Gamma_+}\right|^2\bigg]\cr
&\phantom{{\cal P}_{12}}
       +e^{-\gamma_L (\tau_1+\tau_2)}\bigg[
        2{\gamma\over\Delta\Gamma}
       +{8\over\Delta\Gamma}{\cal R}e(C^*\epsilon_S)
       -16\left|{C\over\Delta\Gamma_+}\right|^2\bigg]\cr
&\phantom{{\cal P}_{12}}
    -{\cal R}e\bigg[\bigg(e^{-\gamma_L\tau_1-\Gamma_-\tau_2}
    +e^{-\Gamma_-\tau_1-\gamma_L\tau_2}\bigg)\ {8C^*\over\Delta\Gamma_+}
                       f_S(-,-)\bigg]\cr
&\phantom{{\cal P}_{12}}
       -2\, {\cal R}e\bigg[e^{-\Gamma_-\tau_1-\Gamma_+\tau_2}\ 
       |f_S(-,-)|^2\bigg]
&(A.9d)\cr
\cr
&{\cal P}_{23}(\tau_1,\tau_2)=\bigg(
        2\, e^{-\gamma_L(\tau_1+\tau_2)}
       -e^{-\gamma_S \tau_1-\gamma_L \tau_2}\bigg)\
        {\gamma\over\Delta\Gamma}\, e^*_L(+,-)\cr
&\phantom{{\cal P}_{12}}
       +e^{-\gamma_L \tau_1-\gamma_S \tau_2}\bigg[
        \left(1+{2(A-\gamma)\over\Delta\Gamma_-}\right)f_S(+,-)
       +{\gamma\over\Delta\Gamma}{\Delta\Gamma_+\over\Delta\Gamma_-}e^*_L(+,-)
       -{2B^*\over\Delta\Gamma_-}f^*_S(+,-)\bigg]\cr
&\phantom{{\cal P}_{12}}
       -e^{-\gamma_L \tau_1-(\Gamma_+ +A-\gamma)\tau_2}\bigg[
        4{C\over\Delta\Gamma_-}
       +{4(A-\gamma)\over\Delta\Gamma_-}\epsilon_S
       +{4\gamma\epsilon_L^*\over\Delta\Gamma_-}
       +{4\gamma C\over\Delta\Gamma\Delta\Gamma_+}
       -{4B^*\epsilon_S^*\over\Delta\Gamma_-}
       +{2iB^*C^*\over\Delta m\Delta\Gamma_+}\bigg]\cr
&\phantom{{\cal P}_{12}}
       +e^{-\gamma_S \tau_1-\Gamma_+ \tau_2}\
        {4\gamma C\over\Delta\Gamma\Delta\Gamma_+}
       +e^{-\gamma_L \tau_1-\Gamma_- \tau_2}\
        {2iB^*C^*\over\Delta m\Delta\Gamma_+}\cr 
&\phantom{{\cal P}_{12}}
       -e^{-\Gamma_-\tau_1-\Gamma_+\tau_2-(A-\gamma)(\tau_1+\tau_2)}\bigg[
       \left(1-{2(A-\gamma)\over\Delta\Gamma_-}\right)f_S(-,-)
       -{2\gamma\over\Delta\Gamma_-}e^*_L(-,-)\cr
&\phantom{{\cal P}_{12}}       
       -{iB^*\over2\Delta m}{\Delta\Gamma_+\over\Delta\Gamma_-}f^*_S(-,-)
       \bigg]
       -e^{-\Gamma_+ (\tau_1+\tau_2)}{iB^*\over\Delta m}f^*_S(-,-)
       +e^{-\Gamma_+\tau_1-\Gamma_-\tau_2}{iB^*\over2\Delta m}f^*_S(-,-)\cr
&&(A.9e)\cr
\cr
&{\cal P}_{33}(\tau_1,\tau_2)=\bigg( 
         e^{-\gamma_S \tau_1-\gamma_L\tau_2}
        +e^{-\gamma_L\tau_1-\gamma_S\tau_2}\bigg)\ e^*_L(+,-)f_S(+,-)\cr
&\phantom{{\cal P}_{12}}        
        -\bigg(e^{-\gamma_S \tau_1-\Gamma_+ \tau_2}
        +e^{-\Gamma_+\tau_1-\gamma_S\tau_2}\bigg)\
        {4C\over\Delta\Gamma_+}f_S(+,-)\cr
&\phantom{{\cal P}_{12}}
        -\bigg(e^{-\gamma_L \tau_1-\Gamma_+ \tau_2}
        +e^{-\Gamma_+\tau_1-\gamma_L\tau_2}\bigg)\
         {4C\over\Delta\Gamma_-}e^*_L(+,-)\cr
&\phantom{{\cal P}_{12}}
        +e^{-(\Gamma_+ +A-\gamma)(\tau_1+\tau_2)}\bigg[
        -{iB^*\over\Delta m}
        -2{iC(\epsilon_S-\epsilon_L^*)\over\Delta m}
        +16{C^2\over|\Delta\Gamma_+|^2}\bigg]\cr
&\phantom{{\cal P}_{12}}
      +\bigg(e^{-\Gamma_+\tau_1-\Gamma_-\tau_2}
      +e^{-\Gamma_-\tau_1-\Gamma_+\tau_2}\bigg)\ e^{-(A-\gamma)(\tau_1+\tau_2)}
      \cr
&\phantom{{\cal P}_{12}}
       \hskip 2cm \times\ \bigg[
        {iB^*\over2\Delta m}
        -e^*_L(-,-)f_S(-,-)
        +i{C(\epsilon_S-\epsilon_L^*)\over\Delta m}
        +8{C^2\over|\Delta \Gamma_+|^2}\bigg]\cr
&&(A.9f)\cr}
$$
\vfill\eject
$$
\eqalignno{
&{\cal P}_{34}(\tau_1,\tau_2)=\ 
         e^{-\gamma_S \tau_1-\gamma_L \tau_2}\ e_L(+,-)f_S(+,-)
        +e^{-\gamma_L \tau_1-\gamma_S \tau_2}\ e^*_L(+,-)f^*_S(+,-)\cr
&\phantom{{\cal P}_{12}}
        -e^{-\gamma_S \tau_1-\Gamma_-\tau_2}\ 
         {4C^*\over\Delta\Gamma_-}f_S(+,-)
        -e^{-\gamma_L \tau_1-\Gamma_-\tau_2}\ 
         {4C^*\over\Delta\Gamma_+}e^*_L(+,-)\cr
&\phantom{{\cal P}_{12}}
        -e^{-\Gamma_+\tau_1-\gamma_S \tau_2}\ 
         {4C\over\Delta\Gamma_+}f^*_S(+,-)
        -e^{-\Gamma_+\tau_1-\gamma_L\tau_2}\ 
         {4C\over\Delta\Gamma_-}e_L(+,-)\cr
&\phantom{{\cal P}_{12}}        
        -e^{-\widetilde{\Gamma}_+\tau_1-\widetilde{\Gamma}_-\tau_2}\bigg[ 1
        +3{|B|^2\over4(\Delta m)^2}
        +2\, {\cal R}e\bigg(\epsilon_S\epsilon_L
        -2{\epsilon_SC^*+\epsilon_LC\over\Delta\Gamma_-}\bigg)
        -24{\cal R}e\bigg({|C|^2\over(\Delta\Gamma_+)^2}\bigg)\bigg]\cr
&\phantom{{\cal P}_{12}}
        +\bigg(2\,e^{-\Gamma_+(\tau_1+\tau_2)}
        +2\, e^{-\Gamma_-(\tau_1+\tau_2)}
        -e^{-\Gamma_-\tau_1-\Gamma_+\tau_2}\bigg)\ {|B|^2\over4(\Delta m)^2}\ .
&(A.9g)
}
$$

\vskip 1cm

{\bf APPENDIX B}
\medskip

We collect here the explicit expressions for the double 
correlations ${\cal G}(f_1,\tau_1;f_2,\tau_2)$. For the observables
having pions in the final states, as in Section 4, 
it is convenient to introduce the following quantities:
$$
\eqalignno{
&R^L_{2\pi}=\left|\epsilon_L+{2C^*\over\Delta\Gamma_-}+Y_{2\pi}\right|^2
         +{\gamma\over\Delta\Gamma}
         -8\left|{C\over\Delta\Gamma_+}\right|^2
         -{4\over\Delta\Gamma}{\cal R}e(C\epsilon_L)\cr
&\eta_{2\pi}=\epsilon_L-{2C^*\over\Delta\Gamma_-}+Y_{2\pi}\ ,&(B.1)\cr
&R^S_{3\pi}=\left|\epsilon_S+{2C\over\Delta\Gamma_-}+Y_{3\pi}\right|^2
          -{\gamma\over\Delta\Gamma}
          -8\left|{C\over\Delta\Gamma_+}\right|^2
          -{4\over\Delta\Gamma}{\cal R}e(C^*\epsilon_S)\cr
&\eta_{3\pi}=\epsilon_S-{2C\over\Delta\Gamma_-}+Y_{3\pi}\ , &(B.2)}
$$
where the indices $2\pi$ and $3\pi$ can assume the values $+-$, $00$
and $+-0$, $000$, respectively, according to the charge of the pions.
Then, using the expressions derived in the previous Appendix, one
explicitly obtains, up to second order in all small parameters: 
$$
\eqalignno{
&{\cal G}(\pi^+\pi^-,\tau_1;\pi^+\pi^-,\tau_2)={|X_{+-}|^4\over 2}\Bigg\{
 e^{-\gamma_S(\tau_1+\tau_2)}\bigg[-2{\gamma\over\Delta\Gamma}
                           +{8\over\Delta\Gamma}{\cal R}e\big(C\epsilon_L\big)
                           -16\left|{C\over\Delta\Gamma_+}\right|^2\bigg]\cr
&\hskip 1cm +\Big(e^{-\gamma_S\tau_1-\gamma_L\tau_2}
+e^{-\gamma_L\tau_1-\gamma_S\tau_2}\Big)\ R^L_{+-}\cr
&\hskip 1cm
-{\cal R}e\bigg[{8C\over\Delta\Gamma_+}\eta_{+-}\Big(
 e^{-\gamma_S\tau_1-\Gamma_+ \tau_2}
+e^{-\Gamma_+ \tau_1-\gamma_S\tau_2}\Big)
+2 |\eta_{+-}|^2\ e^{-\Gamma_+\tau_1-\Gamma_-\tau_2}\bigg]
\Bigg\}
&(B.3a)\cr}
$$
\vfill\eject
$$
\eqalignno{
&{\cal G}(\pi^+\pi^-,\tau_1;2\pi^0,\tau_2)={|X_{+-}|^2|X_{00}|^2\over 2}\Bigg\{
 e^{-\gamma_S(\tau_1+\tau_2)}\bigg[-2{\gamma\over\Delta\Gamma}
                           +{8\over\Delta\Gamma}{\cal R}e\big(C\epsilon_L\big)
                           -16\left|{C\over\Delta\Gamma_+}\right|^2\bigg]\cr
&\hskip .5cm
+e^{-\gamma_S\tau_1-\gamma_L\tau_2}\ R^L_{00}
+e^{-\gamma_L\tau_1-\gamma_S\tau_2}\ R^L_{+-}\cr
&\hskip .5cm
-{\cal R}e\bigg[{8C\over\Delta\Gamma_+}\Big(
\eta_{00}\ e^{-\gamma_S\tau_1-\Gamma_+ \tau_2}
+\eta_{+-}\ e^{-\Gamma_+ \tau_1-\gamma_S \tau_2}\Big) 
+2\, \eta^*_{+-}\, \eta_{00}\ e^{-\Gamma_+\tau_1-\Gamma_-\tau_2}\bigg]\Bigg\}
&(B.3b)\cr
\cr\cr
&{\cal G}(\pi^+\pi^-\pi^0,\tau_1;\pi^+\pi^-\pi^0,\tau_2)=
{|X_{+-0}|^4\over 2}\Bigg\{
e^{-\gamma_L(\tau_1+\tau_2)}\bigg[2{\gamma\over\Delta\Gamma}
                     +{8\over\Delta\Gamma}{\cal R}e\big(C^*\epsilon_S\big)
                     -16\left|{C\over\Delta\Gamma_+}\right|^2\bigg]\cr
&\hskip .5 cm 
+\Big(e^{-\gamma_S\tau_1-\gamma_L\tau_2}
+e^{-\gamma_L\tau_1-\gamma_S\tau_2}\Big)\ R^S_{+-0}\cr
&\hskip .5cm
-{\cal R}e\bigg[{8C^*\over\Delta\Gamma_+}\eta_{+-0}
\Big(e^{-\gamma_L\tau_1-\Gamma_- \tau_2}
+e^{-\Gamma_- \tau_1-\gamma_L\tau_2}\Big)
+2|\eta_{+-0}|^2\ e^{-\Gamma_+\tau_1-\Gamma_-\tau_2}\bigg]\Bigg\}
&(B.3c)\cr
\cr\cr
&{\cal G}(\pi^+\pi^-\pi^0,\tau_1;3\pi^0,\tau_2)=
{|X_{+-0}|^2|X_{000}|^2\over 2}\Bigg\{
e^{-\gamma_L(\tau_1+\tau_2)}\bigg[2{\gamma\over\Delta\Gamma}
                     +{8\over\Delta\Gamma}{\cal R}e\big(C^*\epsilon_S\big)
                     -16\left|{C\over\Delta\Gamma_+}\right|^2\bigg]\cr
&\hskip .5cm
+e^{-\gamma_S\tau_1-\gamma_L\tau_2}\ R^S_{+-0}
+e^{-\gamma_L\tau_1-\gamma_S\tau_2}\ R^S_{000}\cr
&\hskip .5cm
-{\cal R}e\bigg[{8C^*\over\Delta\Gamma_+}\Big(
\eta_{000}\, e^{-\gamma_L\tau_1-\Gamma_- \tau_2}
+\eta_{+-0}\, e^{-\Gamma_- \tau_1-\gamma_L\tau_2}\Big)
+2\, \eta_{000}\, \eta^*_{+-0}\ 
e^{-\Gamma_+\tau_1-\Gamma_-\tau_2}\bigg]\Bigg\}\cr
&&(B.3d)\cr}
$$
The semileptonic correlations (4.13) need first order contributions only:
$$
\eqalignno{
\cr
{\cal G}&(\ell^+,\tau_1;\ell^-,\tau_2)={|{\cal M}|^4\over8}\Bigg\{
          {2\gamma\over\Delta\Gamma}
          \Big(e^{-\gamma_L(\tau_1+\tau_2)}-e^{-\gamma_S(\tau_1+\tau_2)}\Big)
+e^{-\gamma_S\tau_1-\gamma_L\tau_2}+e^{-\gamma_L\tau_1-\gamma_S\tau_2}\cr
&+2\, {\cal R}e\bigg[\epsilon_S-\epsilon_L+x-z
+{4i\beta\over\Delta\Gamma_-}\bigg]
\Big(e^{-\gamma_S\tau_1-\gamma_L\tau_2}
-e^{-\gamma_L\tau_1-\gamma_S\tau_2}\Big)\cr
&+{\cal R}e\bigg[{8C\over\Delta\Gamma_+}\Big(
e^{-\gamma_S\tau_1-\Gamma_+ \tau_2}-e^{-\Gamma_+ \tau_1-\gamma_S \tau_2}
\Big)\bigg]
+{\cal R}e\bigg[{8C\over\Delta\Gamma_-}\Big(
e^{-\gamma_L \tau_1-\Gamma_+ \tau_2}-e^{-\Gamma_+ \tau_1-\gamma_L \tau_2}
\Big)\bigg]\cr
&+e^{-\Gamma(\tau_1+\tau_2)}\bigg[
{4b\over\Delta m}\cos\big(\Delta m(\tau_1+\tau_2)\big)
-\bigg({4b\over\Delta m}-2e^{-(A-\gamma)(\tau_1+\tau_2)}\bigg)
\cos\big(\Delta m(\tau_1-\tau_2)\big)\cr
&+{2(\alpha-a)\over\Delta m}\sin\big(\Delta m(\tau_1+\tau_2)\big)
+4\,{\cal I}m\bigg(\epsilon_L-\epsilon_S+z-x+{4i\beta\over\Delta\Gamma_-}\bigg) 
\, \sin\big(\Delta m(\tau_1-\tau_2)\big)\bigg]\Bigg\}\cr
&&(B.4a)\cr
\cr
{\cal G}&(\ell^+,\tau_1;\ell^+,\tau_2)={|{\cal M}|^4\over8}\Bigg\{
          {2\gamma\over\Delta\Gamma}\bigg(
e^{-\gamma_L(\tau_1+\tau_2)}-e^{-\gamma_S(\tau_1+\tau_2)}\bigg)\cr
&+\bigg[1+2\, {\cal R}e\bigg(\epsilon_S+\epsilon_L-2y
+{4c\over\Delta\Gamma_-}\bigg)\bigg]
\bigg(e^{-\gamma_S\tau_1-\gamma_L\tau_2}
+e^{-\gamma_L\tau_1-\gamma_S\tau_2}\bigg)\cr
&-{\cal R}e\bigg[{8C\over\Delta\Gamma_+}\bigg(
e^{-\gamma_S\tau_1-\Gamma_+ \tau_2}
+e^{-\Gamma_+ \tau_1-\gamma_S \tau_2}\bigg)\bigg]
-{\cal R}e\bigg[{8C\over\Delta\Gamma_-}\bigg(
e^{-\gamma_L \tau_1-\Gamma_+ \tau_2}
+e^{-\Gamma_+ \tau_1-\gamma_L \tau_2}\bigg)\bigg]\cr
&-e^{-\Gamma(\tau_1+\tau_2)}\bigg[
{2(\alpha-a)\over\Delta m}\sin\big(\Delta m(\tau_1+\tau_2)\big)
+{4b\over\Delta m}\cos\big(\Delta m(\tau_1+\tau_2)\big)\cr
&-\bigg({4b\over\Delta m}
-2e^{-(A-\gamma)(\tau_1+\tau_2)}
-4\,{\cal R}e\bigg[\epsilon_L+\epsilon_S-2y-{4c\over\Delta\Gamma_-}\bigg]\bigg)
\cos\big(\Delta m(\tau_1-\tau_2)\big)\bigg]\Bigg\}\cr
&&(B.4b)\cr
\cr
{\cal G}&(\ell^-,\tau_1;\ell^-,\tau_2)={|{\cal M}|^4\over8}\Bigg\{
          {2\gamma\over\Delta\Gamma}\bigg(
e^{-\gamma_L(\tau_1+\tau_2)}-e^{-\gamma_S(\tau_1+\tau_2)}\bigg)\cr
&\bigg[1-2\, 
{\cal R}e\bigg(\epsilon_S+\epsilon_L-2y+{4c\over\Delta\Gamma_-}\bigg)\bigg]
\bigg(e^{-\gamma_S\tau_1-\gamma_L\tau_2}
+e^{-\gamma_L\tau_1-\gamma_S\tau_2}\bigg)\cr
&+{\cal R}e\bigg[{8C\over\Delta\Gamma_+}\bigg(
e^{-\gamma_S\tau_1-\Gamma_+ \tau_2}
+e^{-\Gamma_+ \tau_1-\gamma_S \tau_2}\bigg)\bigg]
+{\cal R}e\bigg[{8C\over\Delta\Gamma_-}\bigg(
e^{-\gamma_L \tau_1-\Gamma_+ \tau_2}
+e^{-\Gamma_+ \tau_1-\gamma_L \tau_2}\bigg)\bigg]\cr
&-e^{-\Gamma(\tau_1+\tau_2)}\bigg[
{2(\alpha-a)\over\Delta m}\sin\big(\Delta m(\tau_1+\tau_2)\big)
+{4b\over\Delta m}\cos\big(\Delta m(\tau_1+\tau_2)\big)\cr
&-\bigg({4b\over\Delta m}
-2e^{-(A-\gamma)(\tau_1+\tau_2)}
+4\,{\cal R}e\bigg[\epsilon_L+\epsilon_S-2y-{4c\over\Delta\Gamma_-}\bigg]\bigg)
\cos\big(\Delta m(\tau_1-\tau_2)\big)\bigg]\Bigg\}\ .\cr
&&(B.4c)\cr
}
$$

\vskip 1cm

{\bf APPENDIX C}
\medskip

As explained in the text, the asymmetry ${\cal A}_{\epsilon^\prime}(\tau)$
in (4.19) allows the determination of real and imaginary parts of the
ratio $\epsilon^\prime/\epsilon$. To first order in this ratio, one finds:
\line{}
$$
{\cal A}_{\epsilon^\prime}(\tau)=3\, 
{\cal R}e\Big({\epsilon^\prime\over\epsilon}\Big)\ {N_1(\tau)\over D(\tau)}
-3\, {\cal I}m\Big({\epsilon^\prime\over\epsilon}\Big)\ 
{N_2(\tau)\over D(\tau)}\ .\eqno(C.1)
$$
The explicit expression of the three coefficients $N_1$, $N_2$ and $D$
is given below.

$$
\eqalignno{
&N_1(\tau)=e^{-\gamma_L\tau}\bigg[ 1 
+4\, {\cal R}e\bigg({C\over\Delta\Gamma_+\, \eta^*_{+-}}\bigg)\bigg]
-e^{-\gamma_S\tau}\bigg[1
+4\,{\cal R}e\bigg({C\over\Delta\Gamma_+\,\eta^*_{+-}}\,
{\gamma_S-\Gamma_-\over\gamma_S+\Gamma_+}\bigg)\bigg]\cr
&\hskip 4cm -4e^{-\Gamma\tau}\,{\cal R}e\bigg(e^{-i\Delta m\tau}
{C\over\Delta\Gamma_+\, \eta^*_{+-}}\,
{2\Gamma\over\gamma_S+\Gamma_+}\bigg)\bigg]\ , &(C.2)\cr
\cr\cr
&N_2(\tau)=4 e^{-\gamma_L\tau}\  
{\cal I}m\bigg({C\over\Delta\Gamma_+\, \eta^*_{+-}}\bigg)
-4 e^{-\gamma_S\tau}\
{\cal I}m\bigg({C\over\Delta\Gamma_+\,\eta^*_{+-}}\,
{\gamma_S-\Gamma_-\over\gamma_S+\Gamma_+}\bigg)\cr
&\hskip 4cm-2e^{-\Gamma\tau}\,{\cal I}m\bigg[e^{-i\Delta m\tau}\bigg(1+
{2C\over\Delta\Gamma_+\, \eta^*_{+-}}\,
{2\Gamma\over\gamma_S+\Gamma_+}\bigg)\bigg]\ , &(C.3)\cr
\cr\cr
&D(\tau)=e^{-\gamma_L\tau}\bigg\{1+{1\over|\eta_{+-}|^2}\bigg[
{\gamma\over\Delta\Gamma}-8\left|{C\over\Delta\Gamma_+}\right|^2-
4\, {\cal R}e\bigg({\epsilon_L C\over\Delta\Gamma}\bigg)\bigg]
+8\, {\cal R}e\bigg({C\over\Delta\Gamma_+\, \eta^*_{+-}}\bigg)\bigg\}\cr
&\ +e^{-\gamma_S\tau}\bigg\{1-{1\over|\eta_{+-}|^2}\bigg[
{\gamma_L\over\gamma_S}\bigg({\gamma\over\Delta\Gamma}
-4\,{\cal R}e\bigg({\epsilon_L C\over\Delta\Gamma}\bigg)+
8\left|{C\over\Delta\Gamma_+}\right|^2\bigg)\bigg]\cr
&+8\, {\cal R}e\bigg({C\over\Delta\Gamma_+\,\eta^*_{+-}}\,
{\gamma_S-\Gamma_-\over\gamma_S+\Gamma_+}\bigg)\bigg\}
-2e^{-\Gamma\tau}\, {\cal R}e\bigg[e^{-i\Delta m\tau}\bigg(1+
{4C\over\Delta\Gamma_+\, \eta^*_{+-}}\,
{2\Gamma\over\gamma_S+\Gamma_+}\bigg)\bigg]\ . \cr
&&(C.4)}
$$

\vskip 1cm

{\bf APPENDIX D}
\medskip

As explained in the text, the radiative decay
$$
\phi\rightarrow\gamma\, K^0\overline{K^0}\ ,\eqno(D.1)
$$ 
could give non-vanishing
contributions to the correlations ${\cal G}(f_1,\tau_1;f_2,\tau_2)$.
These background terms could mimic the effects of the non-standard
parameters $a$, $b$, $c$, $\alpha$, $\beta$, $\gamma$, making more
problematic their experimental determinations.
In this Appendix we collect the explicit expressions
of the contributions $\widetilde{\cal G}$ 
to some of the observables $\cal G$ studied in Section 4,
coming from the background process in $(D.1)$.

In the case of two charged pion final states, one gets, to second order
in all small parameters:
$$
\eqalign{
\widetilde{\cal G}&(\pi^+\pi^-,\tau_1;\pi^+\pi^-,\tau_2)=
{|X_{+-}|^4\over2}\Bigg\{
e^{-\gamma_S(\tau_1+\tau_2)}\bigg\{1-\bigg({\gamma\over\Delta\Gamma}\bigg)^2\cr
&+2\, {\cal R}e\bigg[2\,\bigg(\epsilon_L-{2C^*\over\Delta\Gamma_-}
+Y_{+-}\bigg)
\bigg(\epsilon_S+{2C\over\Delta\Gamma_-}\bigg)-
\bigg(\epsilon_L-{2C^*\over\Delta\Gamma_-}\bigg)^2\bigg]\bigg\}\cr
&+e^{-\gamma_L(\tau_1+\tau_2)}\bigg({\gamma\over\Delta\Gamma}\bigg)^2
-2\,{\cal R}e\bigg[e^{-\Gamma_+(\tau_1+\tau_2)}\bigg(
\epsilon_L-{2C^*\over\Delta\Gamma_-}+Y_{+-}\bigg)^2\bigg]\cr
&+2\,{\cal R}e\bigg[\bigg(e^{-\gamma_S\tau_1-\Gamma_+\tau_2}+
e^{-\Gamma_+\tau_1-\gamma_S\tau_2}\bigg)\bigg(
\epsilon_L-\epsilon_S-2{C+C^*\over\Delta\Gamma_-}\bigg)
\bigg(\epsilon_L-{2C^*\over\Delta\Gamma_-}+Y_{+-}\bigg)\bigg]\Bigg\}\ .\cr}
\eqno(D.2)
$$ 
\line{}
A similar expression holds for the correlation involving three pions in
the final states; explicitly, one finds:
$$
\eqalign{
\widetilde{\cal G}(&\pi^+\pi^-\pi^0,\tau_1;\pi^+\pi^-\pi^0,\tau_2)=
{|X_{+-0}|^4\over2}\Bigg\{
e^{-\gamma_L(\tau_1+\tau_2)}\bigg\{1-\bigg({\gamma\over\Delta\Gamma}\bigg)^2\cr
+&2\, {\cal R}e\bigg[2\,\bigg(\epsilon_S-{2C\over\Delta\Gamma_-}
+Y_{+-0}\bigg)
\bigg(\epsilon_L+{2C^*\over\Delta\Gamma_-}\bigg)-
\bigg(\epsilon_S-{2C\over\Delta\Gamma_-}\bigg)^2\bigg]\bigg\}\cr
+&e^{-\gamma_S(\tau_1+\tau_2)}\bigg({\gamma\over\Delta\Gamma}\bigg)^2
-2\,{\cal R}e\bigg[e^{-\Gamma_-(\tau_1+\tau_2)}\bigg(
\epsilon_S-{2C\over\Delta\Gamma_-}+Y_{+-0}\bigg)^2\bigg]\cr
+&2\,{\cal R}e\bigg[\bigg(e^{-\gamma_L\tau_1-\Gamma_-\tau_2}+
e^{-\Gamma_-\tau_1-\gamma_L\tau_2}\bigg)\bigg(
\epsilon_S-\epsilon_L-2{C+C^*\over\Delta\Gamma_-}\bigg)
\bigg(\epsilon_S-{2C\over\Delta\Gamma_-}+Y_{+-0}\bigg)\bigg]\Bigg\}\ .\cr}
\eqno(D.3)
$$ 
\line{}
Finally, in the case of semileptonic final states, the contribution
coming from the background process $(D.1)$ produces the term:
$$
\eqalign{
\widetilde{\cal G}(\ell^+,\tau_1;\ell^+,\tau_2)=
&{|{\cal M}|^4\over8}\Bigg\{
e^{-\gamma_S(\tau_1+\tau_2)}\bigg[1-{2\gamma\over\Delta\Gamma}
+4\,{\cal R}e\bigg(\epsilon_S+x-y+{2C\over\Delta\Gamma_-}
\bigg)\bigg]\cr
&+e^{-\gamma_L(\tau_1+\tau_2)}\bigg[1+{2\gamma\over\Delta\Gamma}
+4\,{\cal R}e\bigg(\epsilon_L-x-y+{2C\over\Delta\Gamma_+}
\bigg)\bigg]\cr
&-2\,{\cal R}e\bigg\{e^{-\Gamma_+(\tau_1+\tau_2)}\bigg[
1-4\Big({\cal R}e(y)+i{\cal I}m(x)\Big)
+2\bigg(\epsilon_L-{2C^*\over\Delta\Gamma_-}\bigg)\cr
&\hskip 5.5cm +2\bigg(\epsilon_S^*-{2C^*\over\Delta\Gamma_+}\bigg)
+2i{B\over2\Delta m}\bigg]\bigg\}\cr
&-2\,{\cal R}e\bigg[\bigg(e^{-\gamma_S\tau_1-\Gamma_+\tau_2}+
e^{-\Gamma_+\tau_1-\gamma_S\tau_2}\bigg) \bigg(\epsilon_S-\epsilon_L
+2{C+C^*\over\Delta\Gamma_-}\bigg)\bigg]\cr
&-2\,{\cal R}e\bigg[\bigg(e^{-\gamma_L\tau_1-\Gamma_+\tau_2}+
e^{-\Gamma_+\tau_1-\gamma_L\tau_2}\bigg) \bigg(\epsilon_L^*-\epsilon_S^*
+2{C+C^*\over\Delta\Gamma_+}\bigg)\bigg]\cr
&-2\bigg(e^{-\Gamma_+\tau_1-\Gamma_-\tau_2}+
e^{-\Gamma_-\tau_1-\Gamma_+\tau_2}\bigg)\, 
{\cal I}m\bigg({B\over2\Delta m}\bigg)
\Bigg\}\ .}
\eqno(D.4)
$$

\vfill\eject

\centerline{\bf REFERENCES}
\bigskip

\item{1.} J. Ellis, J.S. Hagelin, D.V. Nanopoulos and M. Srednicki,
Nucl. Phys. {\bf B241} (1984) 381
\smallskip
\item{2.} J. Ellis, N.E. Mavromatos and D.V. Nanopoulos, Phys. Lett.
{\bf B293} (1992) 142
\smallskip
\item{3.} J. Ellis, N.E. Mavromatos and D.V. Nanopoulos, Int. J. Mod. Phys.
{\bf A11} (1996) 1489
\smallskip
\item{4.} J. Ellis, J.L. Lopez, N.E. Mavromatos and D.V. Nanopoulos, 
Phys. Rev. D {\bf 53} (1996) 3846
\smallskip
\item{5.} P. Huet and M.E. Peskin, Nucl. Phys. {\bf B434} (1995) 3
\smallskip
\item{6.} S. Hawking, Comm. Math. Phys. {\bf 87} (1983) 395
\smallskip
\item{7.} M. Srednicki, Nucl. Phys. {\bf B410} (1993) 143
\smallskip
\item{8.} CPLEAR Collaboration, J. Ellis, N.E. Mavromatos and 
D.V. Nanopoulos, Phys. Lett. {\bf B364} (1995) 239
\smallskip
\item{9.} R. Alicki and K. Lendi, {\it Quantum Dynamical Semigroups and 
Applications}, Lect. Notes Phys. {\bf 286}, (Springer-Verlag, Berlin, 1987)
\smallskip
\item{10.} H. Spohn, Rev. Mod. Phys. {\bf 53} (1980) 569
\smallskip
\item{11.} L. Fonda, G.C. Ghirardi and A. Rimini, Rep. Prog. Phys.
{\bf 41} (1978) 587
\smallskip
\item{12.} V. Gorini, A. Kossakowski and E.C.G. Sudarshan,
J. Math. Phys. {\bf 17} (1976) 821
\smallskip
\item{13.} G. Lindblad,  Comm. Math. Phys. {\bf 48} (1976) 119
\smallskip
\item{14.} E.B. Davies, {\it Quantum Theory of Open Systems}, (Academic Press,
New York, 1976)
\smallskip
\item{15.} V. Gorini, A. Frigerio, M. Verri, A. Kossakowski and
E.C.G. Surdarshan, Rep. Math. Phys. {\bf 13} (1978) 149 
\smallskip
\item{16.} F. Benatti and R. Floreanini, Phys. Lett. {\bf B389} (1996) 100
\smallskip
\item{17.} F. Benatti and R. Floreanini, Nucl. Phys. {\bf B488} (1997) 335
\smallskip
\item{18.} F. Benatti and R. Floreanini, Testing complete positivity,
Trieste-preprint, 1996
\smallskip
\item{19.} F. Benatti and R. Floreanini, Experimental limits on
complete positivity from the $K-\overline{K}$ system, Phys. Lett. B, to appear
\smallskip
\item{20.} T.D. Lee and C.S. Wu, Ann. Rev. Nucl. Sci. {\bf 16} (1966) 511
\smallskip
\item{21.} I. Dunietz, J. Hauser and J.L. Rosner, Phys. Rev. D {\bf 35}
(1987) 2166
\smallskip
\item{22.} J. Bernabeu, F.J. Botella and J. Roldan, Phys. Lett. {\bf B211}
(1988) 226
\smallskip
\item{23.} C.D. Buchanan, R. Cousins, C. Dib, R.D. Peccei and
J. Quackenbush, Phys. Rev. D {\bf 45} (1992) 4088
\smallskip
\item{24.} M. Fukawa, Y. Fukushima, K. Hirata, Y. Kawashima,
S.K. Kim, T. Ohshima, J. Shirai, T. Taniguchi, T. Tanimori, S.L. Olsen,
K. Ueno, F. Sannes, S. Schnetzer, K. Kinoshita, Prog. Theor. Phys. Suppl.
{\bf 119} (1995) 1
\smallskip
\item{25.} {\it The Second Da$\,\mit\Phi$ne Physics Handbook}, 
L. Maiani, G. Pancheri and N. Paver, eds., (INFN, Frascati, 1995)
\smallskip
\item{26.} L. Wolfenstein, CP violation, in {\it Fundamental Symmetries},
P. Bloch, P. Pavlopoulos and R. Klapish, eds. (Plenum, New York, 1986)
\smallskip
\item{27.} N.W. Tanner and R.H. Dalitz, Ann. of Phys. {\bf 171} (1986) 463
\smallskip
\item{28.} G. D'Ambrosio, G. Isidori and A. Pugliese, $CP$ and $CPT$
measurements at Da$\Phi$ne, in Ref.[25]
\smallskip
\item{29.} G. D'Ambrosio and N. Paver, Phys. Rev. D {\bf 49} (1994) 4560
\smallskip
\item{30.} N. Paver and Riazuddin, Phys. Lett. {\bf B246} (1990) 240
\smallskip
\item{31.} N. Brown and F.E. Close, Scalar mesons and kaons in $\phi$
radiative decays and their implications for studies of $CP$ violation at 
Da$\Phi$ne, in Ref.[25]
\smallskip
\item{32.} N.N. Achasov, V.V. Gubin and V.I. Shevchenko,
Phys. Atom. Nucl. {\bf 60} (1997) 81
\smallskip
\item{33.} D. Cocolicchio, G. Fogli, M. Lusignoli and A. Pugliese,
Phys. Lett. {\bf B238} (1990) 417

\bye